\documentclass[a4paper,12pt]{article}
\pdfoutput=1 

\usepackage{jheppub,amsmath} 
\usepackage{graphicx}
\usepackage{dcolumn}
\usepackage{bm}
\usepackage{amsmath,amsfonts,amssymb}
\usepackage{slashed, comment}

\DeclareMathOperator{\sech}{sech}

\usepackage{braket,xcolor}
\usepackage{verbatim}
\usepackage{caption}
\usepackage{subcaption}
\usepackage{multirow}
\usepackage{amsfonts}
\usepackage[utf8]{inputenc}
\usepackage{hyperref}
\usepackage{braket}
\usepackage[T1]{fontenc} 

\title{\boldmath Krylov complexity in large $q$ and double-scaled SYK model}

\author[a]{Budhaditya Bhattacharjee,}
\author[b]{Pratik Nandy,}
\author[a]{and Tanay Pathak}
\affiliation[a]{Centre for High Energy Physics, Indian Institute of Science,\\ C.V. Raman Avenue, Bangalore 560012, India}
\affiliation[b]{Center for Gravitational Physics and Quantum Information,\\
Yukawa Institute for Theoretical Physics, Kyoto University,\\
Kitashirakawa Oiwakecho, Sakyo-ku, Kyoto 606-8502, Japan}

\emailAdd{budhadityab@iisc.ac.in}
\emailAdd{pratik@yukawa.kyoto-u.ac.jp}
\emailAdd{tanaypathak@iisc.ac.in}

\abstract{Considering the large $q$ expansion of the Sachdev-Ye-Kitaev (SYK) model in the two-stage limit, we compute the Lanczos coefficients, Krylov complexity, and the higher Krylov cumulants in subleading order, along with the $t/q$ effects. The Krylov complexity naturally describes the ``size'' of the distribution while the higher cumulants encode richer information. We further consider the double-scaled limit of SYK$_q$ at infinite temperature, where $q \sim \sqrt{N}$. In such a limit, we find that the scrambling time shrinks to zero, and the Lanczos coefficients diverge. The growth of Krylov complexity appears to be ``hyperfast'', which is previously conjectured to be associated with scrambling in de Sitter space.}

\begin{document} 
\maketitle
\flushbottom

\section{Introduction}
\label{sec:intro}
Understanding quantum chaos has been a long-standing problem in theoretical physics. It also serves as a powerful microscope for probing the features of holographic dualities. Classically, chaotic dynamics are fairly well understood. It is based on the phase space trajectories under infinitesimal perturbations in the initial condition, whose exponential deviation is often called the ``butterfly effect'' \cite{Roberts:2014ifa, Roberts:2016wdl}. In quantum mechanics, there is a large ambiguity in the definition of chaos. This is because trajectories are ill-defined objects in the realm of quantum dynamics. One of the well-accepted definitions of quantum chaos comes from the level statistics of the eigenspectrum of the quantum system. The distribution followed by the eigenvalues reflects level crossing or level repulsion in the system, which is believed to be an underlying signature of chaotic dynamics \cite{PhysRevE.81.036206}. In the last few years, many indirect probes of scrambling and quantum chaos have been proposed. These include operator distribution \cite{Roberts:2014isa, Roberts:2018mnp, Qi:2018bje, Lensky:2020ubw, Schuster:2021uvg}, out-of-time-ordered-correlators (OTOCs) \cite{Rozenbaum:2016mmv, Hashimoto:2017oit, Nahum:2017yvy, Xu:2019lhc, Zhou:2021syv, Gu:2021xaj}, and Krylov complexity \cite{Parker:2018yvk, Barbon:2019wsy, Rabinovici:2020ryf, Bhattacharjee:2022vlt}. These probes have been tested against quantum mechanical realizations of known classical chaotic systems, purely quantum systems, and black holes in holography. In this work, we study one such probe, Krylov complexity (K-complexity in short), and its various cousins in Sachdev-Ye-Kitaev (SYK) model \cite{PhysRevLett.70.3339, Kittu}. As a probe of quantum-chaotic dynamics, it has been studied extensively over the past few years. Applications of Krylov complexity extend from a few body quantum systems to field theories \cite{Barbon:2019wsy, Bhattacharjee:2022vlt, Avdoshkin:2019trj, Dymarsky:2019elm, Jian:2020qpp, Rabinovici:2020ryf, Cao:2020zls, Dymarsky:2021bjq, Kar:2021nbm, Caputa:2021sib, Kim:2021okd, Magan:2020iac, Caputa:2021ori,  Bhattacharya:2022gbz, Rabinovici:2022beu, Muck:2022xfc, Liu:2022god, Patramanis:2021lkx, Caputa:2022eye, Rabinovici:2021qqt, Fan:2022xaa, Fan:2022mdw, Trigueros:2021rwj,  Hornedal:2022pkc, Balasubramanian:2022tpr, Bhattacharjee:2022qjw, Heveling:2022hth, Caputa:2022yju, Balasubramanian:2022dnj, Afrasiar:2022efk, Guo:2022hui}.

The main focus of our attention is the Sachdev-Ye-Kitaev (SYK) model, originally proposed as a model with all-to-all random fermion interactions in $0+1$ dimensions by Sachdev and Ye \cite{PhysRevLett.70.3339}. This model was then studied by Kitaev as a simple model for holography, where he demonstrated that it serves as a holographic dual to a black hole in AdS$_{2}$ geometry \cite{Kittu}. It is a simple model of $N$ fermions with $q$-body random interactions, which is exactly solvable in the large $N$ and large $q$ limit \cite{Maldacena:2016hyu}.\footnote{This model is also exactly solvable in the $q=2$ limit. However, in this limit, the Hamiltonian is just a random mass-like free fermion \cite{Magan:2015yoa, Anninos:2016szt} and the model becomes integrable \cite{Maldacena:2016hyu, Gross:2016kjj}.} It is also known as a maximal scrambler \cite{Sekino:2008he, Lashkari:2011yi}, in the sense that it saturates the Maldacena-Shenker-Stanford bound on chaos \cite{Maldacena:2015waa} and exhibits the random-matrix like statistics in the late-time behavior of the spectral form factor \cite{PhysRevE.55.4067, Garcia-Garcia:2016mno, Cotler:2016fpe, Saad:2018bqo}. Past studies have been extended to various notions of operator growth with recent explorations of Krylov complexity in this model \cite{Parker:2018yvk, Rabinovici:2020ryf}. Especially, the leading order large $q$ results have been considered in \cite{Parker:2018yvk}, where exact analytic results have been obtained. We extend the study to the next order in the large $q$ expansion. We employ the Lanczos algorithm to auto-correlation function for the model where correction terms up to order $\mathcal{O}(1/q^{2})$ are considered. We discuss the effect of the $\mathcal{O}(1/q^{2})$ term on the Lanczos coefficients and the Krylov basis wavefunctions. Utilizing the distribution, we compute the first three cumulants of the distribution and discuss their interpretation.

Next, we turn our attention to the operator growth and scrambling evolved by the SYK Hamiltonian with a modified variance of the distribution for the random coupling. We achieve this modified distribution by re-scaling and taking the double-scaled limit in the usual SYK model \cite{Berkooz:2018jqr, Berkooz:2020uly}. We denote it by $\mathrm{DSSYK}_{\infty}$ where the suffix ``$\infty$'' indicates that we are in at infinite temperature (defined through the Boltzmann distribution). The growth of the size of an operator in this system is then measured by the Krylov complexity. This is a systematic approach for examining the operator growth, which nicely complements the framework of the epidemic model \cite{Qi:2018bje} studied in \cite{Lin:2022nss, Susskind:2022bia, Rahman:2022jsf}. Further, given the probability distribution, the higher cumulants can also be computed exactly. The complexity appears to be exponential, and the Lyapunov coefficient diverges as $q \rightarrow \infty$. Analogously, this implies the vanishing scrambling time in the large $q$ limit. In previous literature, it is termed hyperfast scrambling and conjectured to be associated with the operator growth in de Sitter (dS) space \cite{Susskind:2021esx, Susskind:2022dfz, Lin:2022nss}. Several recent studies have also explored holographic complexity and scrambling in dS space. See \cite{Chapman:2021eyy, Jorstad:2022mls} for more details. 

The paper is organized as follows. In section \ref{kcomsec}, we briefly review the Krylov construction and the cumulants of its distribution functions. Section \ref{syklargeq} starts with the setup of the SYK Hamiltonian, with the $1/q$ order correction of the auto-correlation function, and the derivation of Lanczos coefficients and various Krylov cumulants. This is followed by the $1/q^2$ order correction with the resulting Lanczos coefficients and Krylov cumulants. We compare both the results in finite and large $q$ limit. In section \ref{dShyperfast}, we study a rescaled SYK with a particular double-scaled limit and focus on the hyperfast scrambling, which is conjectured to be associated with operator growth in dS space. We finally conclude the paper with some future remarks on the possible future direction in continuation to this present work. 

At the final stages of this work, two papers \cite{Susskind:2022bia, Rahman:2022jsf} appear discussing the hyperfast scrambling in DSSYK$_{\infty}$ using the framework of the epidemic model. Here we exploit the construction of Krylov complexity for the same, and our results are consistent with their findings.

\section{Lanczos coefficients and Krylov cumulants}\label{kcomsec}
We start with a brief review of operator growth and Krylov complexity (K-complexity). Under the time-evolution governed by some time-independent Hamiltonian $H$, an initial operator (may be properly normalized) $\mathcal{O}_{0}$ evolves as
\begin{equation}
\mathcal{O}(t)=e^{i H t / \hbar} \,\mathcal{O}_0\, e^{-i H t / \hbar}\,.
\end{equation}
The growth can be understood by evaluating the nested commutators obtained by the Baker-Campbell-Hausdorff (BCH) expansion. However, the nested commutators do not form an orthonormal basis. One efficient way to form such an orthonormal basis is to apply the Gram-Schmidt orthonormalization-like procedure, usually known as the Lanczos algorithm \cite{viswanath1994recursion}. The resulting basis is known as the Krylov basis, and the set of normalization coefficients, known as the Lanczos coefficients, correctly captures the growth of such operators. The time-evolved operator is expressed on the Krylov basis as \cite{Parker:2018yvk}
\begin{align}
|\mathcal{O}(t))=\sum_{n=0}^{\mathcal{K}-1} i^n \varphi_n (t) | \mathcal{O}_n)\,,
\end{align}
where $\mathcal{K}$ is the Krylov dimension. The $\varphi_{n}$'s appearing in the above equation is the Krylov basis functions, and they follow the following recursive differential equation \cite{Parker:2018yvk}
\begin{align} 
    \partial_t \varphi_n(t) = b_n \varphi_{n-1}(t) -  b_{n+1} \varphi_{n+1}(t) \,, \label{recde}
\end{align}
where $b_n$'s are the Lanczos coefficients. The $|\varphi_{n}(t)|^2$ defines the probability with $\sum_{n} |\varphi_{n}(t)|^{2} = 1$ for all time. The zeroth-order basis function $\varphi_0(t)$ relates the two-point auto-correlation function
\begin{align}
    C(t) \equiv \varphi_0(t) = (\mathcal{O}_0 | \mathcal{O}(t))\,, \label{autoq0}
\end{align}
where the infinite-temperature inner-product is defined as $(A|B)_{\beta = 0} = \frac{1}{D} \mathrm{Tr}[A^{\dagger} B]$. For the case of SYK, we have $D = 2^{N/2}$. The auto-correlation function \eqref{autoq0} can be expanded in a Taylor series. The coefficients of the series are known as moments. The moments can be calculated as \cite{viswanath1994recursion, Parker:2018yvk}
\begin{align}
    C(-it) = \sum_{n=0}^{\infty} m_{2n} \,\frac{t^{2n}}{(2n)!}\,, ~~~~~ m_{2n} = (-1)^{n} \frac{d^{2n}}{d t^{2n}} C(t)\bigg|_{t = 0}\,. \label{momentsdef}
\end{align}
with $m_0 = 1$. If the initial operator is Hermitian, all the odd moments vanish. It is important to note that these moments can also be obtained as \cite{viswanath1994recursion, Parker:2018yvk}
\begin{align}
    m_{2n} = \frac{1}{2\pi} \int_{-\infty}^{\infty} \mathrm{d} \omega\, \omega^{2n} \,\Phi (\omega)\,, ~~~~~~ \Phi (\omega) = \int_{-\infty}^{\infty} \mathrm{d} t \,e^{- i \omega t} C(t)\,,
\end{align}
where $\Phi (\omega)$ is known as the spectral density. As long as the auto-correlation function is given, the spectral density can be straightforwardly computed, and the (even) moments can be obtained. However, it is interesting to understand how far the converse statement is true, namely, given the set of moments, can one construct the spectral density and, thereby, the auto-correlation function? Is the construction unique? This is a well-known Hamburger moment problem\footnote{There exists other moments problem namely the Stieltjes moment problem defined on $[0, \infty)$ and the Hausdorff moment problem defined on $[0, 1]$. The latter is always determinate \cite{chihara}.} defined on $\mathbb{R} \equiv (-\infty, \infty)$ \cite{parkerthesis}. If the spectral density can be obtained uniquely, then the moment problem is called \emph{determinate}; otherwise, it is referred to as \emph{indeterminate}. We will come back to these questions in later sections.

Once moments are given, one can directly apply to the following iterative algorithm \cite{viswanath1994recursion, Parker:2018yvk}
\begin{align}
    b_n &= \sqrt{Q_{2n}^{(n)}}\,, ~~~~ Q_{2k}^{(m)} = \frac{Q_{2k}^{(m-1)}}{b_{m-1}^2} - \frac{Q_{2k-2}^{(m-2)}}{b_{m-2}^2}\,, \nonumber \\
    Q_{2k}^{(0)} &= m_{2 k}\,, ~~~~ b_{-1} = b_{0} := 1\,, ~~~~ Q_{2k}^{(-1)} := 0\,. \label{mombn}
\end{align}
to find the Lanczos coefficients $b_n$. One then performs the iterative recursion \eqref{recde} to obtain the Krylov basis wavefunctions $\varphi_n (t)$'s.

Now, we define the cumulants of the distribution $\varphi_n (t)$. The average position of the probability distribution, called the Krylov complexity ($C_K$) and the (normalized) variance, called the Krylov variance ($\delta_K$) as follows \cite{Parker:2018yvk, Barbon:2019wsy, Caputa:2021ori}
\begin{align}
    C_K (t) &= \sum_{n} n |\varphi_n (t)|^2\,, \label{kcom} \\ 
    \delta_K(t) &= \frac{\sum n^2 |\varphi_n (t)|^2 - (\sum n |\varphi_n (t)|^2)^2}{(\sum n |\varphi_n (t)|^2)^2}\,. \label{kvar}
\end{align}
These quantities capture qualitative information about the distribution function. We further define the third cumulant, the Krylov skewness as\footnote{From now, we usually omit the prefix ``K'' from K-complexity, K-variance, and K-skewness and just refer to them as complexity, variance, and skewness, respectively.}
\begin{align}
    s_K (t) = \frac{\sum n^3 |\varphi_n (t)|^2 - 3 \sum n |\varphi_n (t)|^2 \big(\sum n^2 |\varphi_n (t)|^2 - (\sum n |\varphi_n (t)|^2)^2 \big) - \big(\sum n |\varphi_n (t)|^2\big)^3}{\big(\sum n^2 |\varphi_n (t)|^2 - (\sum n |\varphi_n (t)|^2)^2 \big)^{3/2}}\,,
\end{align}
which encodes much richer information. Given the analytic forms of the distribution, in principle, the cumulants can be evaluated exactly. This can be efficiently done by considering the following cumulant generating function \cite{Qi:2018bje, Roberts:2018mnp, Jian:2020qpp}
\begin{align}
    \log\braket{e^{\lambda \hat{K}}} = \log(O(t)|e^{\lambda \hat{K}}|O(t)) =   \log\Big(\sum_{n} e^{\lambda n} \, |\varphi_n (t)|^2 \Big)\,.
\end{align}
One can now take the $n$-derivative to compute the $n$-th cumulant of the distribution 
\begin{align}
    k_n = \partial_{\lambda}^k\log\braket{e^{\lambda \hat{K}}}|_{\lambda = 0}\,. \label{kcum}
\end{align}
For example, it is evident that the K-complexity is the first cumulant $k_1$ of the $\braket{e^{\lambda \hat{K}}}$ operator. Similarly, it is easy to see the higher cumulants provide the variance and the skewness. Although The higher cumulants encode more information, in many cases, they can be expressed in terms of the lower cumulants. In this article, we only focus on the first three cumulants of the distribution.

\section{SYK in the large $q$ expansion}\label{syklargeq}

The well-known Sachdev-Ye-Kitaev (SYK) model is a $0+1$-dimensional fermionic model with $N \gg 1$ fermions, where each fermion is coupled randomly with others. The $q$-body (we take $q$ even) interaction Hamiltonian is given by \cite{Maldacena:2016hyu}
\begin{align}
    H = i^{q/2} \sum_{1 \leq i_1 < \cdots <  i_q \leq N} j_{i_1 \cdots i_q} \, \psi_{i_1} \cdots \psi_{i_q}\,, \label{sykh1old}
\end{align}
where $j_{i_1 \cdots i_q}$ are random couplings, drawn from some Gaussian ensemble with zero mean $\braket{j_{i_1 \cdots i_q}} = 0$ and the variance given by
\begin{align}
    \braket{j^2_{i_1 \cdots i_q}} = \frac{(q-1)!}{N^{q-1}} J^2 = 2^{q-1} \frac{(q-1)!}{q N^{q-1}} \mathcal{J}^2 \,.
\end{align}
where $\mathcal{J}^2 = 2^{1-q} q J^2$. The factor $i$ is chosen to make the Hamiltonian Hermitian, and the factor $N$ in the variance makes the model interesting in the large $N$ limit. The constant $J$ is a dimensionful parameter and sets the energy scale of the Hamiltonian. The fermions satisfy the anti-commutation relation
\begin{align}
    \{\psi_a, \psi_b\} = \delta_{ab}\,. \label{anc}
\end{align}
It is convenient to rescale the field as $\chi_a = \sqrt{2} \psi_a$. The re-scaled fields satisfy the anti-commutation relation $\{\chi_a, \chi_b\} = 2\delta_{ab}$. In other words, this redefines the Hamiltonian as
\begin{align}
    H = i^{q/2} \sum_{1 \leq i_1 < \cdots <  i_q \leq N} \mathbf{j}_{i_1 \cdots i_q} \, \chi_{i_1} \cdots \chi_{i_q}\,, \label{sykh1}
\end{align}
with the variance
\begin{align}
    \braket{\mathbf{j}^2_{i_1 \cdots i_q}} = 2^{-q} \braket{j^2_{i_1 \cdots i_q}} =  \frac{(q-1)!}{2 q N^{q-1}} \mathcal{J}^2\,.
\end{align}
For $N$ fermions, the dimension of the Hilbert space is $2^{N/2}$. The model simplifies dramatically in the limit when the number of fermions, $N$, is large. In this limit, only the melonic diagrams contribute to the Schwinger-Dyson equation and the model self averages, i.e., one obtains identical results for the correlation functions for any randomly chosen couplings. Moreover, two specific cases are exactly solvable, one is the $q=2$ limit, which is integrable \cite{Maldacena:2016hyu, Gross:2016kjj} and another is the large $q$ limit \cite{Maldacena:2016hyu}. All $q \geq 4$ are chaotic and share similar properties. The large $q$ limit is rather interesting, and one can take this limit in the following two ways.
\begin{enumerate}
    \item The two-stage limit:  In this case, we first take the $N \rightarrow \infty$ limit with $q$ fixed and then take the $q \rightarrow \infty$ limit. This is the standard procedure that is followed in various places \cite{Gu:2021xaj, Maldacena:2016hyu, Cotler:2016fpe, Jia:2018ccl, Maldacena:2018lmt, Choi:2019bmd, Jiang:2019pam}. Our results with the $1/q^{2}$ term have been derived in this limit.

    \item The double-scaled limit: In this case, we simultaneously take $N \rightarrow \infty$, $q \rightarrow \infty$ keeping $\lambda=q^{2}/N \rightarrow$ fixed.\footnote{In the gravity picture, a dual $\lambda$ can be defined in terms of Planck scale ($L_m$) and string scale ($L_s$). It has a close similarity with 't Hooft coupling. See \cite{Susskind:2022bia} for more details.} This is also been used at various places \cite{Cotler:2016fpe, Berkooz:2018jqr, Berkooz:2022fso, Berkooz:2020uly, Susskind:2021esx, Susskind:2022dfz, Lin:2022nss, Lin:2022rbf, Khramtsov:2020bvs}. This is more general than the two-stage limit where the two-stage limit is supposed to recover at fixed $q$ and $N \rightarrow \infty$ i.e., at $\lambda = 0$ limit \cite{Berkooz:2018jqr, Rahman:2022jsf}.\footnote{To preserve the ``locality'', one requires either fixed $q$ or the infinite limit scaling $q \sim N^{a}$, with $a < 1/2$ \cite{Cotler:2016fpe, Gharibyan:2018jrp}. The scaling $q \sim N^{1/2}$ marks the transition from the semicircle density of states to the Gaussian ones as we decrease $a$ \cite{French:1971akf}. Hence, the double scaling limit with $q \sim N^{a}$ with $a > 1/2$ marks the non-locality \cite{erdHos2014phase}. Also see \cite{Garcia-Garcia:2017pzl, Garcia-Garcia:2018fns}.} 
\end{enumerate}
In the next section, we will systematically study the operator growth in the $ 1/q$ expansion of the SYK model.

\subsection{Including $1/q$ correction}
We start with the normalized initial operator $O(0) \equiv \sqrt{2} \psi_1 (0) = \chi_1(0)$. The auto-correlation function is given by the following two-point function
\begin{align}
     \mathcal{C} (\tau) = (O(\tau) |O(0))_{\beta}\,,
\end{align}
with respect to the finite-temperature inner product with temperature $1/\beta$.
Assuming $q$-large, we expand the above auto-correlation in a series of $1/q$ as \cite{Maldacena:2016hyu}
\begin{align}
    \mathcal{C} (\tau) &= 1 + \frac{1}{q}\, g(\tau)  + \cdots\,. \label{autoq}
\end{align}
Here we only keep the sub-leading term, where the function $g(\tau)$ obeys the following Liouville differential equation \cite{Parker:2018yvk}
\begin{align}
    \partial_{\tau}^2 g &= - 2 \mathcal{J}^2 e^{g}\,, ~~~~~~~~ \mathcal{J}^2 = 2^{1-q} q J^2\,, \label{ode1}
\end{align}
with $\mathcal{J}$ remains fixed. In this article, we focus on the infinite-temperature limit. With the boundary conditions $g(0) = 0$, $g'(0) = 0$, the solution of the above equation is given by
\begin{align}
    g(t) &= 2 \ln (\sech \mathcal{J} t )\,,
\end{align}
where $\tau = i t$. The auto-correlation function \eqref{autoq} is then expanded in a Taylor series, and the corresponding moments are evaluated by using \eqref{momentsdef}. They are given by
\begin{align}
    m_{2n} = \frac{1}{q}  \mathcal{J}^{2n} T_{n-1} + O(1/q^2)\,, ~~~~~~ n \geq 1\,, \label{mom1}
\end{align}
where $\{T_{n-1}\}_{n=1}^{\infty} = \{1, 2, 16, 272, 7936, \cdots \}$ are the Tangent numbers. For large $n$, the moments do not increase quite rapidly as $q \rightarrow \infty$. Especially, we can see
\begin{align}
    \lim_{N \rightarrow \infty} \sum_{n = 1}^N m_{2n}^{-1/2n}  = \frac{1}{\mathcal{J}} \lim_{N \rightarrow \infty} \sum_{n = 1}^N q^{1/2n} T_{n-1}^{-1/2n} =  \infty\,.
\end{align}
Applying the iterative algorithm \eqref{mombn}, we find the Lanczos coefficients as
\begin{align}
b_n =
  \begin{cases}
    \mathcal{J}\sqrt{2/q} + O(1/q) \,,      & ~~n = 1\,,\\
    \mathcal{J}\sqrt{n(n-1)} + O(1/q) \,,   & ~~n > 1\,.  \label{bnleading}
  \end{cases}
\end{align}
Alternatively, this implies
\begin{align}
    \lim_{N \rightarrow \infty} \sum_{n = 1}^N \frac{1}{b_n} = \frac{1}{\mathcal{J}}\sqrt{\frac{q}{2}} +  \frac{1}{\mathcal{J}} \lim_{N \rightarrow \infty} \sum_{n = 2}^N \frac{1}{\sqrt{n(n-1)}} = \infty\,.
\end{align}
Hence, the Carleman’s condition \cite{carleman} is satisfied and the Hamburger moment problem is determinate \cite{chihara}. In other words, the determinate nature of the moment problem guarantees that the Lanczos coefficients are bounded (i.e., $b_n$ cannot grow more than linearly in $n$).\footnote{In section \ref{dShyperfast}, we will see that the unbounded (divergent) growth of $b_n$ fails to satisfy Carleman’s condition, and the determinate nature can be questioned.} The Krylov basis functions are obtained by solving the recursive differential equation \eqref{recde}. They are given by
\begin{align}
\varphi_n(t) =
  \begin{cases}
    1 + (2/q) \ln (\sech \mathcal{J} t) + O(1/q^2) \,,      & ~~n = 0\,,\\
    \sqrt{2/n\, q}\, \tanh^n (\mathcal{J} t) + O(1/q^2) \,,   & ~~n \geq 1\,. \label{diss}
  \end{cases}
\end{align}
The probability amplitude is $|\varphi_n(t)|^2$, and the total probability sums up to unity, i.e., $\sum_{n=0}^{\infty}|\varphi_n(t)|^2 = 1$, which can be straightforwardly checked.  
Using the form of $\varphi_n(t)$, we compute the complexity and the variance
\begin{align}
    C_K (t) = \frac{2}{q} \sinh^2 (\mathcal{J} t) + O(1/q^2)\,, ~~~~~  \delta_k (t) = \frac{q}{2} \coth^2 (\mathcal{J} t) - 1 + O(1/q^2)\,. \label{cc} 
\end{align}
The leading-order behavior of complexity is dominated by $1/q$, whereas the variance is proportional to $q$ to the leading order. This was reported in \cite{Roberts:2018mnp} in the context of the size of the operator. 
We further compute the skewness
\begin{align}
    s_{K} (t) = \sqrt{2 q} \coth(2 \mathcal{J} t) -\frac{3 \tanh (\mathcal{J} t) \text{sech}^2(\mathcal{J} t)}{\sqrt{2 q}} + O(1/q^{3/2})\,. \label{skr}
\end{align}
The leading expression of skewness is proportional to $\sqrt{q}$. However, we should mention that we only trust the results in the leading order, as we have taken the auto-correlation function up to the $O(1/q)$ order. In the next section, we will follow up on the $O(1/q^2)$ correction to the auto-correlation function, and consequently, we will be able to comment on the subleading contribution of Lanczos coefficients and the associative quantities like complexity, variance, and skewness.

The previous results do not take into account that the $t$ scales with $q$ in large $q$ limit. To account for the $t/q$-correction, we consider the following probability distribution \cite{Roberts:2018mnp}
\begin{align}
    P_n(t) = \frac{\Gamma (n+2/q)}{\Gamma (n+1) \,\Gamma(2/q)}\, \frac{\tanh^{2 n}(\mathcal{J} t)}{\cosh^{4/q}(\mathcal{J} t)}\,, \label{prob1}
\end{align}
which yields the same result as \eqref{cc}  and \eqref{skr} for complexity, variance, and skewness in the leading expression.\footnote{In the leading order in $1/q$ expansion, the factor in front of $\tanh^{2 n}(\mathcal{J} t)$ becomes $2/nq$, which relates \eqref{diss} as $P_n(t) = |\varphi_n(t)|^2$.} This distribution sums up to unity i.e., $\sum_{n=0}^{\infty} P_n(t) = 1$ for all time $t$. According to \cite{Roberts:2018mnp}, this distribution defines the ``size'' of the operator.

As a final note, it is also sometimes useful to consider the Krylov entropy (K-entropy) for the distribution \eqref{diss}. The K-entropy is defined as \cite{Barbon:2019wsy}
\begin{align}
    E_{K} (t) = - \sum_{n} | \varphi_{n}(t) | \ln | \varphi_n (t) |^2\,. \label{ent}
\end{align}
The analytical expression for $E_{K}(t)$ at order $\mathcal{O}(1/q)$ is straightforward to obtain using the expression for $\varphi_{n}(t)$. The result is
\begin{align}
    E_{K} (t) = &-\frac{2}{q}\text{Li}_{1}^{(1,0)}\left(\tanh^2 t\right)-\frac{4}{q}\sinh^2 t \log (\tanh t) -\frac{4}{q}\log(\sech t)\left(1+\log\left(\frac{q}{2}\right)\right)\,, \label{cent}
 \end{align}
where $\text{Li}_{1}^{(1,0)}(z) \equiv \partial_{s}\text{Li}_{s}(z)|_{s \to 1}$. The initial growth of the entropy is linear (see Fig.\,\ref{entplots} in the later section) which is consistent with the exponential growth of the complexity.

\subsection{Including $1/q^2$ correction}

\begin{figure}[t]
		\centering
		\begin{subfigure}[b]{0.48\textwidth}
		\centering
		\includegraphics[width=\textwidth]{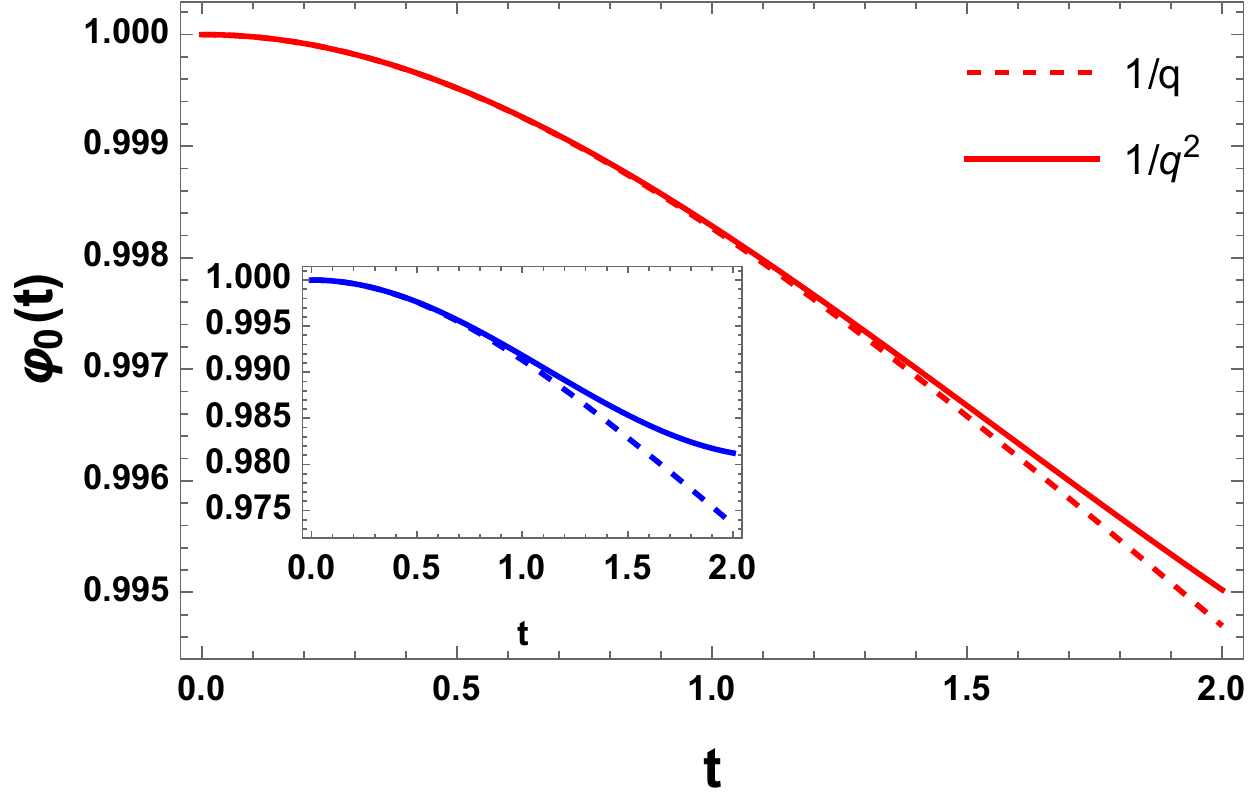}
			\caption{}
			\label{fig:psi0}
		\end{subfigure}
		\hfill
		\begin{subfigure}[b]{0.48\textwidth}
		\centering
		\includegraphics[width=\textwidth]{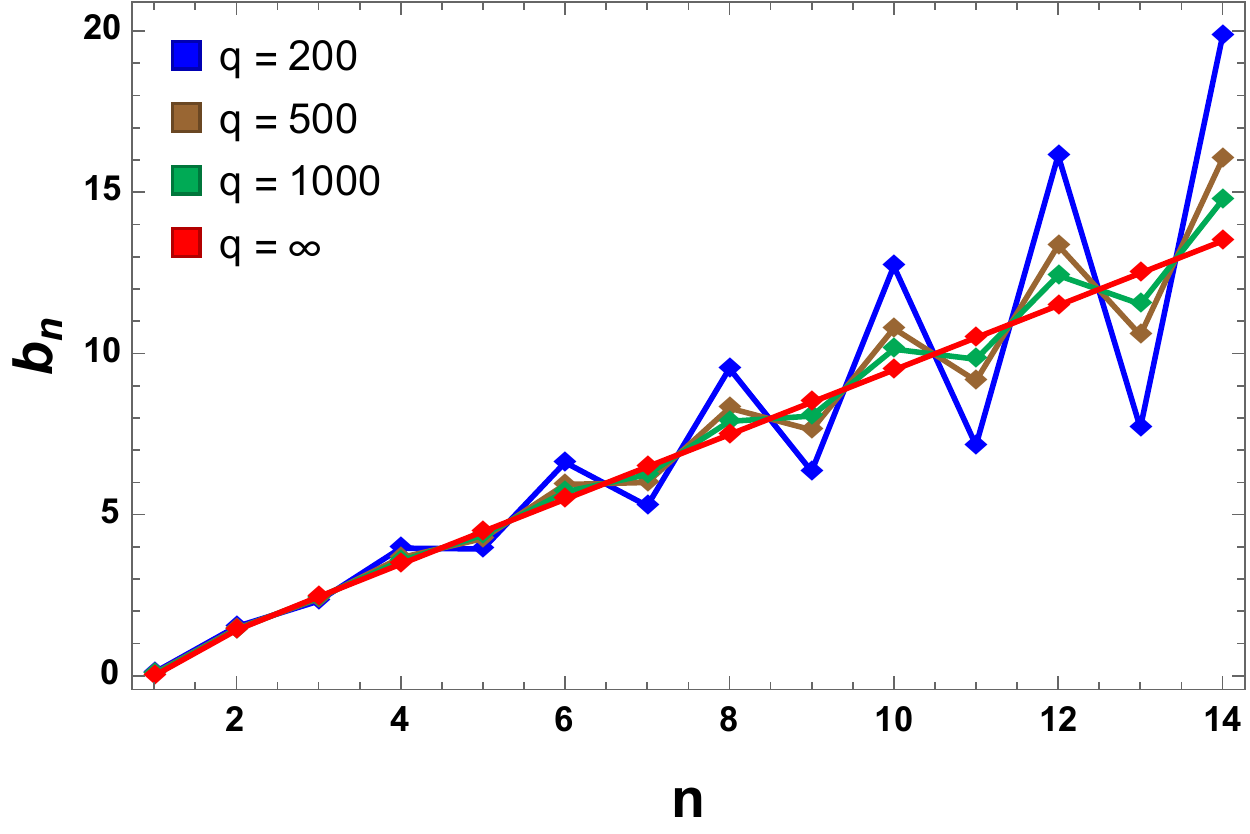}
			\caption{}
			\label{fig:bn}
		\end{subfigure}
		\caption{(a) Comparison of the auto-correlation function $\varphi_0 (t)$ up to the $O(1/q)$ and $O(1/q^2)$ for $q=500$ (red). Inset shows the comparison for $q=100$. The dashed line gives the analytic result \eqref{diss}. (b) The Lanczos coefficients $b_n$ in the large $q$ limit of the SYK model, keeping the $O(1/q)$ correction to the $b_n$s. We set $\mathcal{J} = 1$.}
		\label{kcplots1}
\end{figure}

Next, we add the sub-subleading $1/q^2$ correction to the auto-correlation function. We write
\begin{align}
    \mathcal{C} (\tau) &= 1 + \frac{1}{q}\, g(\tau)  + \frac{1}{q^2}\, h(\tau) + \cdots \,,\label{Green1}
\end{align}
where both the terms $g(\tau)$ and $h(\tau)$ satisfy the following differential equations \cite{Tarnopolsky:2018env}
\begin{align}
    \partial_{\tau}^2 g &= 2 \mathcal{J}^2 e^{g}\,, \\
    \partial_{\tau}^2 h &= 2 \mathcal{J}^2 e^{g} h + \frac{1}{2} \partial_{\tau}^3 (g \star g) - 2 \mathcal{J}^2 e^g \bigg(g + \frac{g^2}{2} \bigg)\,, \label{tareq1}
\end{align}
where $\mathcal{J}^2 = 2^{1-q} q J^2$ and the convolution is defined as
\begin{align}
    \frac{\pi v}{\beta} \frac{1}{2} \partial_{x}^3 (g \star g) = 2 \partial_x \bigg[g(\tau) \bigg\{\cot\bigg(\frac{\pi v}{\beta} + x\bigg) - \cot\bigg(\frac{\pi v}{\beta} - x\bigg) \bigg\}\bigg] - 4\,. \label{conv1}
\end{align}
Here we express the variable $x$ as $x = \frac{\pi v}{2} - \frac{\pi v \tau}{\beta}$, with $0 \leq v \leq 1$. The high temperature (weak coupling) implies $v \rightarrow 0$ while the low temperature (strong coupling) sets $v \rightarrow 1$ \cite{Choi:2019bmd}. In terms of the $\tau$ variable, the convolution reads
\begin{align}
    \partial_{\tau}^3 (g \star g) =  \frac{4 \pi v}{\beta}  \partial_{\tau} \bigg[g(\tau) \bigg\{\cot\bigg(\frac{\pi v}{2} + \frac{\pi v}{\beta} (1 - \tau) \bigg) - \cot\bigg(\frac{\pi v}{2} - \frac{\pi v}{\beta} (1 + \tau) \bigg) \bigg\}\bigg] - 4\,. \label{conv2}
\end{align}
The Eq.\eqref{tareq1} can be solved analytically and more interestingly in a closed form. With the boundary condition $g(\tau=0) = g(\tau=\beta) = 0$ and $h(\tau=0) = h(\tau=\beta) = 0$, we have the solution \cite{Tarnopolsky:2018env}
\begin{align}
    g(\tau) &= 2 \ln \bigg[\cos\Big(\frac{\pi v}{2}\Big)\sec\Big(\frac{\pi v}{2} - \frac{\pi v \tau}{\beta} \Big)\bigg]\,,~~~~~~~~~~ \beta \mathcal{J} = \pi v \sec \Big(\frac{\pi v}{2}\Big)\,,\\
    h(\tau) &= \frac{1}{2} g^2(\tau) - 2 \ell(\tau) - 4 \bigg[ \tan \Big(\frac{\pi v}{2} - \frac{\pi v \tau}{\beta} \Big) \int_{0}^{\frac{\pi v}{2} - \frac{\pi v \tau}{\beta}} dy \,\ell(y) + 1\bigg]  \nonumber \\
    &+ 4 \,\frac{1 + \Big(\frac{\pi v}{2} - \frac{\pi v \tau}{\beta} \Big) \tan \Big(\frac{\pi v}{2} - \frac{\pi v \tau}{\beta} \Big)}{1 + \frac{\pi v}{2} \cos\big(\frac{\pi v}{2}\big)} \bigg[ \tan \frac{\pi v}{2} \int_{0}^{\pi v/2} dy \,\ell(y) + 1\bigg] \,, \label{htau}
\end{align}
where $\ell(y) \equiv g(y) - e^{-g(y)} \mathrm{Li}_2(1-e^{g(y)})$. The detailed computation is given in Appendix \ref{appauto}. This gives the full auto-correlation function up to $O(1/q^2)$. In Fig.\,\ref{fig:psi0}, we compare the auto-correlation function at the $1/q$ and $1/q^2$ order for $q = 500$ and $q = 100$ (inset). 

\subsection{Moments and Lanczos coefficients}

Performing the derivatives according to \eqref{momentsdef}, we evaluate the corresponding moments. The non-zero moments are given by
\begin{align}
     m_{2n} = \frac{1}{q}  \mathcal{J}^{2n} T_{n-1} + \frac{1}{q^2}  \mathcal{J}^{2n} t_{n-1} + O(1/q^3)\,, ~~~~~~ n \geq 1\,,
\end{align}
where, as before, $\{T_{n-1}\}_{n=1}^{\infty} = \{1, 2, 16, 272, 7936, \cdots \}$ are the tangent numbers, and $\{t_{n-1}\}_{n=1}^{\infty} = \{0, 64, 368, 11440, 406864, 22368256, 1640452864, \cdots \}$. In Appendix \ref{appa}, we list up to $m_{30}$. However, there appears to be no well-known sequence for these sub-leading terms to us. We now execute the iterative (and tedious) algorithm \eqref{mombn} to compute the Lanczos coefficients. They are given by
\begin{align}
b_n =
  \begin{cases}
    \mathcal{J}\sqrt{2/q} + O(1/q) \,,      & ~~n = 1\,,\\
    \mathcal{J}\sqrt{n(n-1)} + \mathfrak{b}_n \, \mathcal{J}/q + O(1/q^2) \,,   & ~~n > 1\,.  \label{bnsubleading}
  \end{cases}
\end{align}
where $\mathfrak{b_n}$'s are given as $\{\mathfrak{b}_{n}\}_{n=2}^{\infty} = \{\frac{31}{\sqrt{2}}, - \frac{65}{\sqrt{6}}, \frac{343}{\sqrt{12}}, - \frac{8677}{18 \sqrt{20}}, \frac{74987}{60\sqrt{30}}, \cdots \}$. Computing higher coefficients is more time-consuming. They are shown in Fig.\,\ref{fig:bn} for different values of $q$ (see Appendix \ref{appa} for a list of the first 14 Lanczos coefficients). The red curve shows the leading contribution $b_n \propto \sqrt{n(n-1)}$ which is independent of $q$. From the expressions of $b_n$'s, we see that $q$ starts to contribute in the subleading order, which is expected. A potentially interesting point to note is that the subleading terms contribute positively to leading terms for the even moments, while the subleading terms negatively contribute to leading terms for the odd moments. These odd and even oscillations are more apparent for small $q$-limit. The odd and even Lanczos coefficients grow very differently.\footnote{Another example of an oscillatory growth pattern of $b_n$'s was previously observed for an artificial auto-correlation function \cite{Avdoshkin:2019trj}. We thank Anatoly Dymarsky for pointing this out to us.} In particular, the even Lanczos coefficients show super-linear growth while the odd coefficients are sublinear. As is demonstrated in Section 4, the oscillatory Lanczos coefficients actually lead to Krylov cumulants that grow faster than their $\mathcal{O}(1/q)$ versions. This tells us that simply having oscillatory Lanczos coefficients is not sufficient to force the K-cumulants to decrease as compared to the ones corresponding to non-oscillatory versions. The exact nature of the oscillations also plays a significant role. As soon as $q$ increases, they start falling on a single line. In the limit, $q \rightarrow \infty$, all the subleading terms vanish, and \eqref{bnleading} is recovered.

\begin{figure}[t]
		\centering
		\begin{subfigure}[b]{0.48\textwidth}
		\centering
		\includegraphics[width=\textwidth]{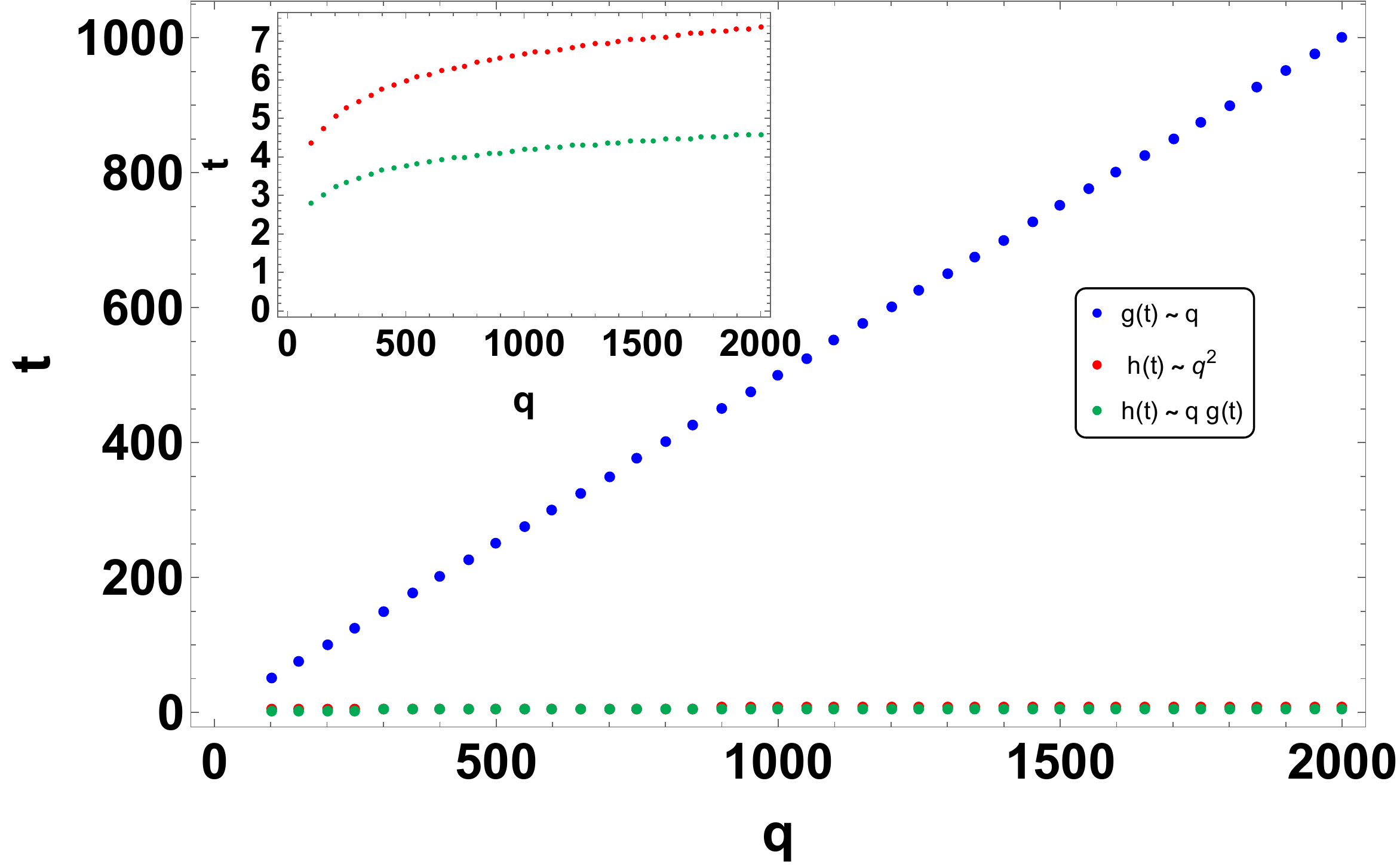}
			\caption{Various time scales.}
			\label{fig:tvsq}
		\end{subfigure}
		\hfill
		\begin{subfigure}[b]{0.48\textwidth}
		\centering
		\includegraphics[width=\textwidth]{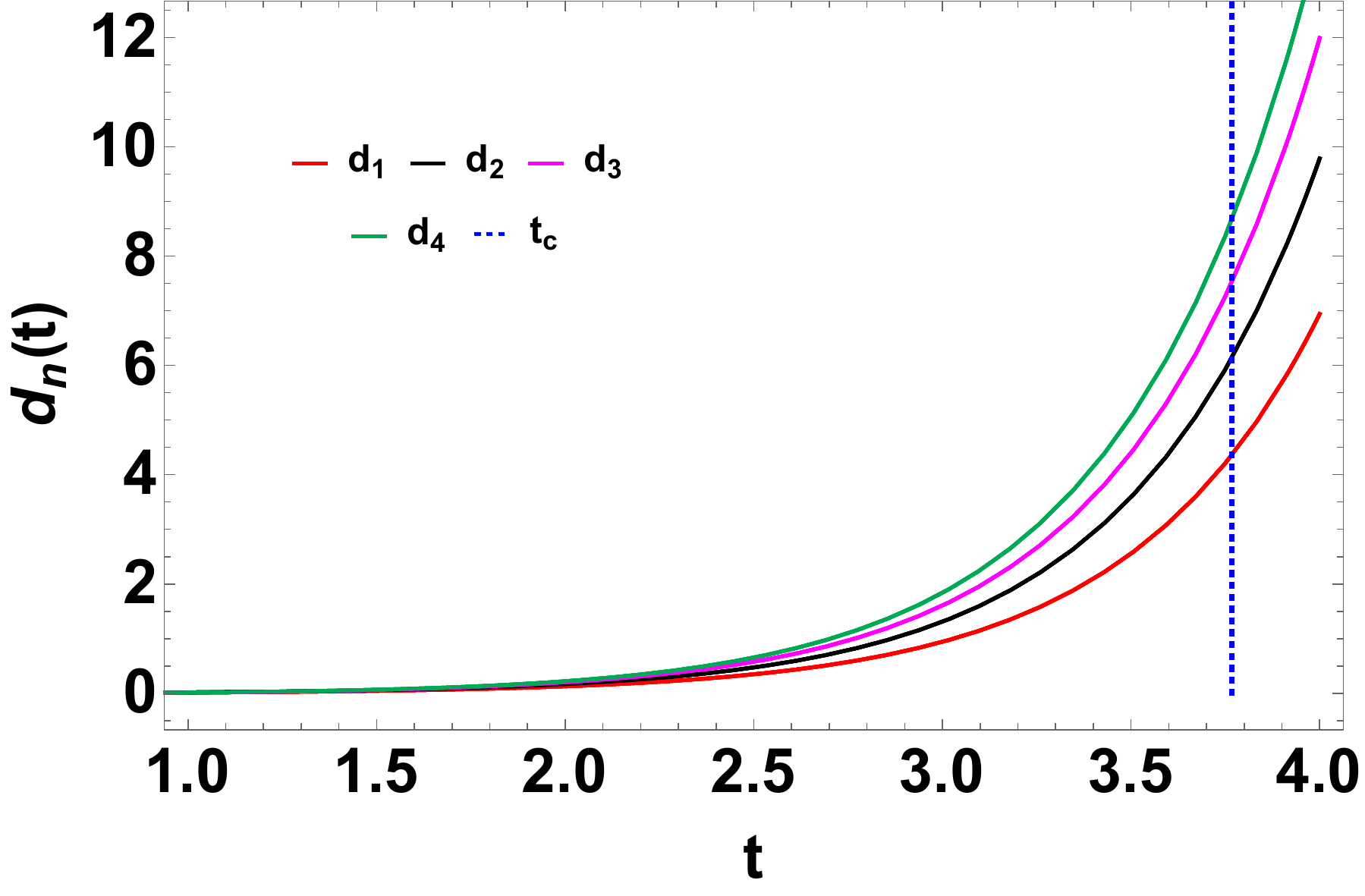}
			\caption{For $q = 100$.}
			\label{fig:wvdiff}
		\end{subfigure}
		\caption{(a) The different time-scale up to where the perturbation theory \eqref{Green1} is valid. The blue, red and the green dots indicate $|g(t)| \sim q$, $|h(t)| \sim q^2$ and $|h(t)| \sim q \,|g(t)|$ respectively. To compute the $t_c$, we choose the last one, namely $|h(t_c)| \sim q \,|g(t_c)|$. (b) The difference $d_{n} \equiv |\varphi^{(1/q^2)}_{n}(t) - \varphi^{(1/q)}_{n}(t)|$ is plotted, between the Krylov wavefunctions at the $1/q$ and $1/q^2$ orders for $q = 100$. The plots correspond to $d_{1}, d_{2}, d_{3}$ and $d_{4}$. As is seen, $d_{n}$ increases significantly (odd wavefunctions increase negatively while the even wavefunctions increase positively) well before the cutoff time.}
		\label{ph}
\end{figure}

In Fig.\,\ref{fig:wvdiff}, we compare the few Krylov wavefunctions at the $1/q$ and $1/q^2$ order for $q = 100$. We find that much before cutoff time $t_c$ (see Fig.\,\ref{fig:tvsq}), there is a significant deviation of the $1/q^2$ results from the known $1/q$ result. We notice that for even $n$, the $1/q^2$ order wavefunctions increases with $t$ as compared to the $1/q$ wavefunctions and vice versa.

\section{Results of Krylov cumulants}
In this section, we discuss the properties of the Krylov cumulants obtained from the $\mathcal{O}(1/q^{2})$ correction to the SYK Greens' function \eqref{Green1}. From \eqref{Green1} and the Lanczos coefficients \eqref{bnsubleading}, we analytically obtain the Krylov basis wavefunctions $\varphi_{n}(t)$, truncating up to $\mathcal{O}(1/q^2)$. Treating their squared sum $\sum_{n} |\varphi_n (t)|^{2}$ as the probability distribution, we study the first few cumulants of the same. These cumulants are the K-complexity, K-variance, and K-skewness, as described in Section \ref{kcomsec}. We can then compare the numerical results obtained with the analytic results, which are obtained considering up to the $\mathcal{O}(1/q)$ correction \eqref{autoq}.\footnote{We should mention that the truncation in $n$ deviates the system away from the real SYK system at all times. It does indeed work as long as the number of wavefunctions considered describes the system completely. However, it mimics the real SYK at early times. We emphasize that the time up to which the truncated system (comprising of the first $13$ wavefunctions) captures the full system (as determined by evaluating the probability $\sum_{n = 1}^{13}|\varphi_{n}(t)|^2$) is greater than the time at which the perturbation series fails $t_{c}$ and conclusions are valid since we only make our observations up to $t_{c}$. We thank the anonymous referee to point out this subtle issue to us.}

\subsection{Krylov cumulants with truncation in $q$} \label{wtr}
As evident from the previous section, we use the auto-correlation function truncated up to $\mathcal{O}(1/q^2)$ to obtain the Lanczos coefficients up to $\mathcal{O}(1/q)$. From this point, we can proceed in two different ways. First, we calculate wave functions $\varphi_{n}$ with proper truncation so that the final results for Krylov complexity, variance, skewness, and entropy are obtained up to $\mathcal{O}(1/q^2)$. For the results, we have considered the first $13$ wavefunctions for our computations. There are two effects that come into play here. The first effect arises due to finite $n$. Since we are taking a finite number of wavefunctions to construct the probability distribution, after some cutoff time (dependent on the details of the system), the squared sum becomes less than unity. This can be understood in the operator spreading picture as well. From that perspective, the finite $\varphi_n (t)$'s fall short of capturing the Krylov basis into which the operator spreads. In other words, after some finite $t$, the $(n + 1)^{\mathrm{th}}$ Krylov basis vector becomes significant, and so the $(n + 1)^{\mathrm{th}}$ wavefunction needs to be included to describe the operator spreading.

\begin{figure}[t]
		\centering
		\begin{subfigure}[b]{0.48\textwidth}
		\centering
		\includegraphics[width=\textwidth]{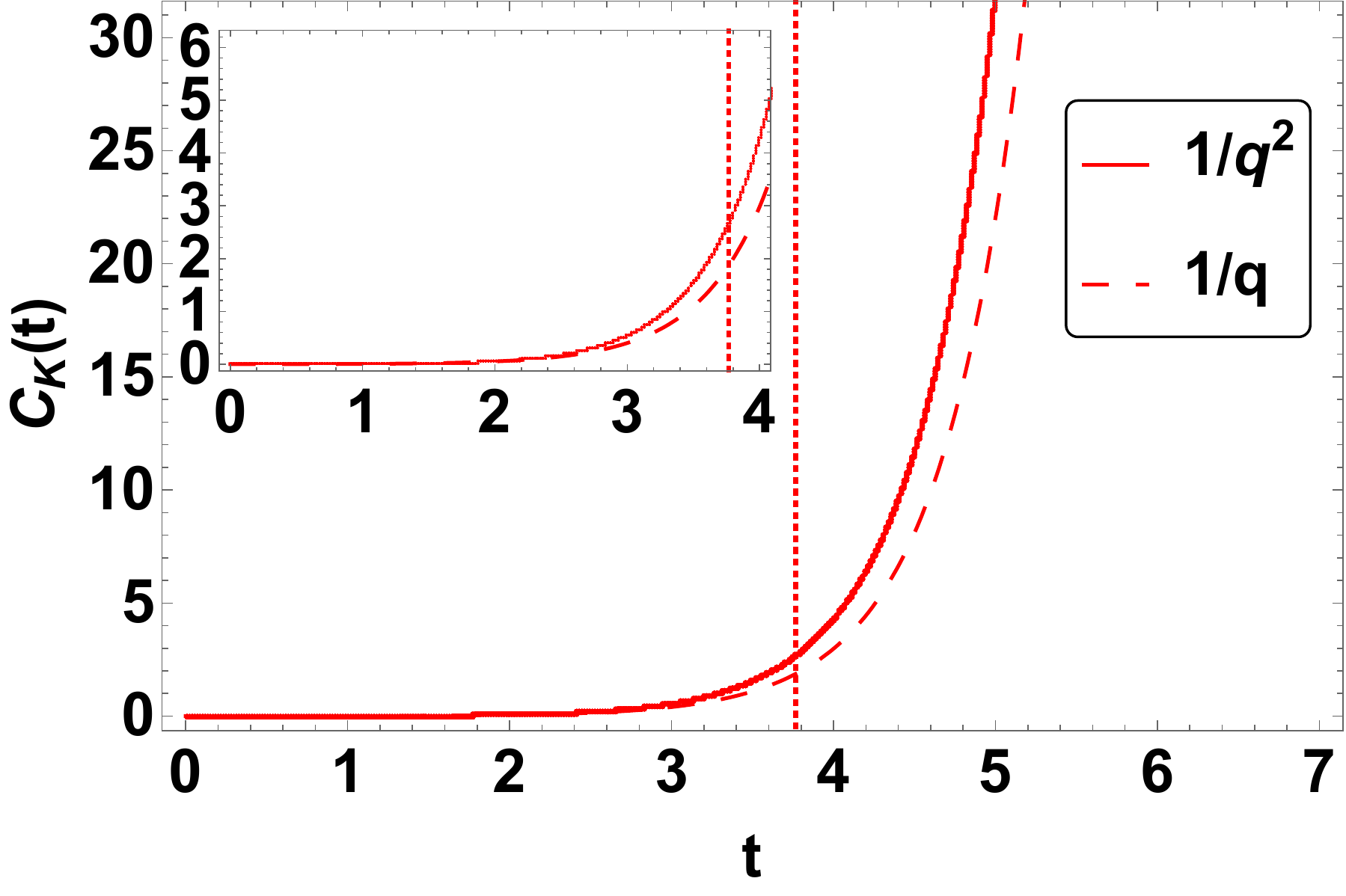}
			\caption{For $q = 500$.}
			\label{fig:kc3}
		\end{subfigure}
		\hfill
		\begin{subfigure}[b]{0.48\textwidth}
		\centering
		\includegraphics[width=\textwidth]{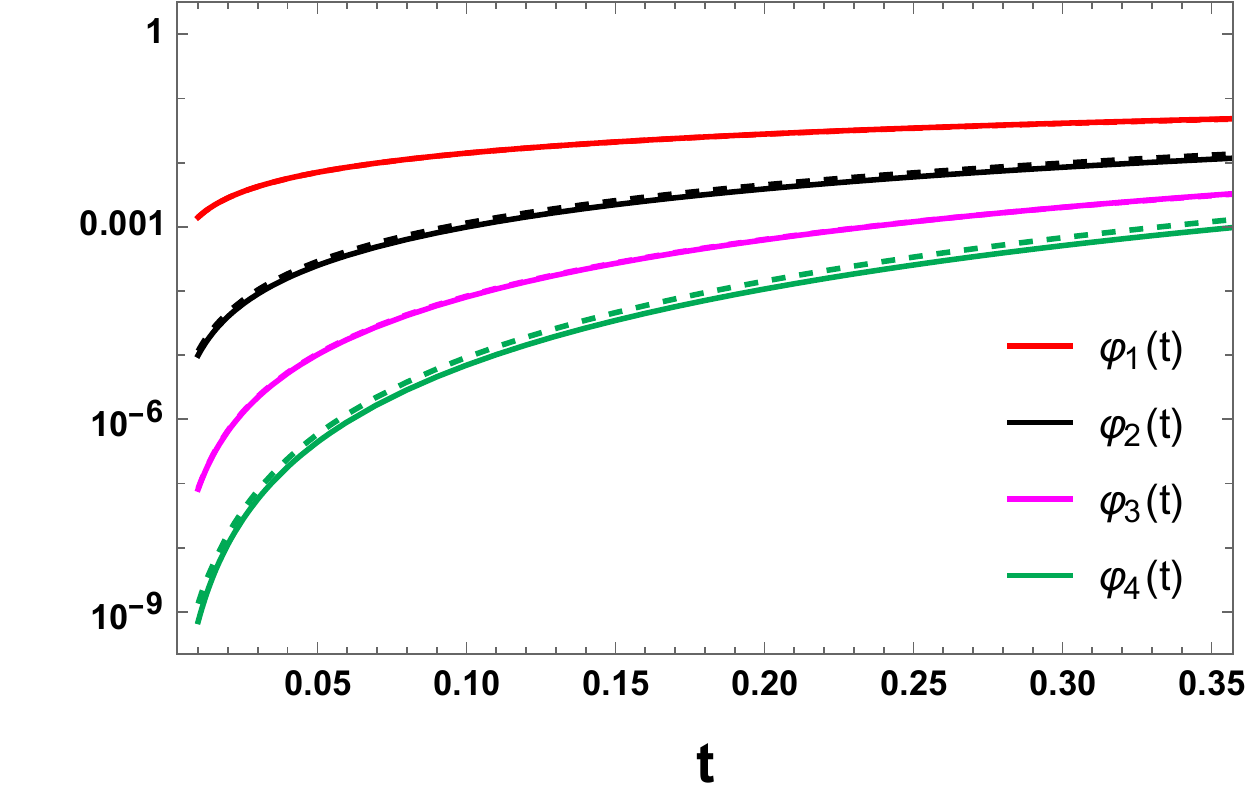}
			\caption{For $q = 100$.}
			\label{fig:phin}
		\end{subfigure}
		\caption{(a) The behavior of truncated K-complexity up to $\mathcal{O}(1/q^{2})$ for $q = 500$. The dashed line is the analytic result \eqref{cc}. The inset shows the early time behavior. The dotted lines represent the time $t_{c}$ up to which our results are reliable. (b) Various $\varphi_n(t)$ is shown for $q=100$. These are obtained by taking the auto-correlation function \eqref{Green1} and the Lanczos coefficients \eqref{bnsubleading}. These wave functions are evaluated without any truncation in $q$. The dashed line indicates the analytic result \eqref{diss}. We observe that the matching of odd wavefunctions is better than the even wavefunctions.}
		\label{kcplots}
\end{figure}

The second effect arises due to the perturbative series \eqref{Green1} failing at large enough times. This can be interpreted as a finite $t/q$ scaling effect \cite{Roberts:2018mnp}. Simply put, it is a finite time $t_c$ after which one or more terms in the series \eqref{Green1} become $\mathcal{O}(1)$ numbers, and hence the expansion fails. However, a shorter time at which the series perturbative approximation fails is when the $\mathcal{O}(1/q)$ term and $\mathcal{O}(1/q^2)$ terms become comparable. Fig.\,\ref{fig:tvsq} shows the comparison between the time scales with the variation of $q$. We treat this timescale as $t_c$ and demonstrate that this is indeed the shorter one of the other timescales obtained by comparing the $\mathcal{O}(1/q)$ and $\mathcal{O}(1/q^2)$ terms to $\mathcal{O}(1)$. This time $t_c$ is the point up to which our results for the moments are reliable. In other words, the $t/q$ effects become dominant after $t = t_c$. We numerically find that for large $q$, $t_c$ increases polynomially in $q$, with a small exponent. The finite $n$ effects do not play a significant role up to $t = t_c$ for the $\varphi_n (t)$'s considered in our computations. Happily, even at times $t < t_c$, the effect of the $\mathcal{O}(1/q^2)$ term in the perturbation series becomes evident, and we observe a significant deviation from the $\mathcal{O}(1/q)$ results, for all the wavefunctions (see Fig.\,\ref{fig:wvdiff}). The effect is more pronounced for higher values of $q$ (and correspondingly higher $t_c$).
\begin{figure}[t]
		\centering
		\begin{subfigure}[b]{0.48\textwidth}
		\centering
		\includegraphics[width=\textwidth]{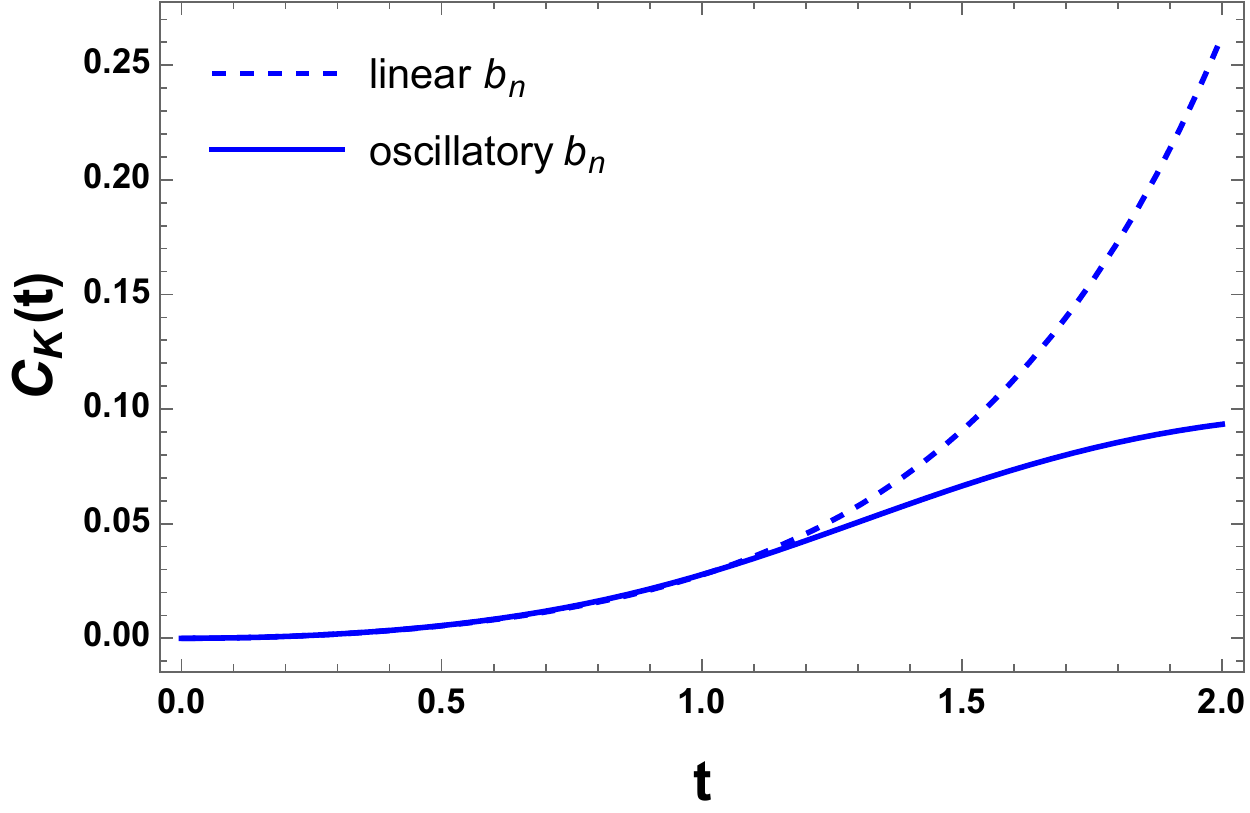}
			\caption{For $q = 100$.}
			\label{fig:var3}
		\end{subfigure}
		\hfill
		\begin{subfigure}[b]{0.48\textwidth}
		\centering
		\includegraphics[width=\textwidth]{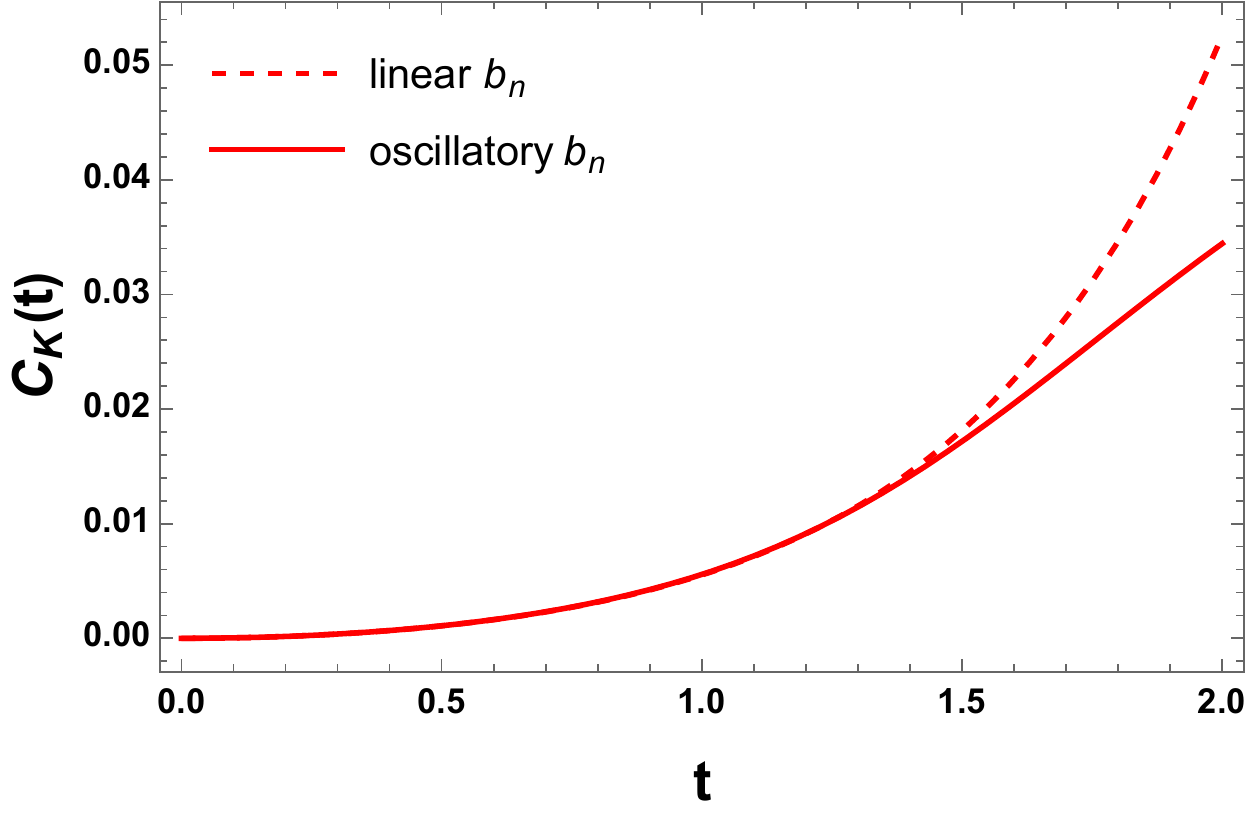}
			\caption{For $q = 500$.}
			\label{fig:var4}
		\end{subfigure}
		\caption{The behavior of K-complexity  is shown for (a) $q=100$ and (b) $q = 500$ respectively, with the Lanczos coefficients \eqref{bnsubleading}. The dashed line demonstrates the analytic result \eqref{cc}.}
		\label{complots}
\end{figure}

\begin{figure}[t]
		\centering
		\begin{subfigure}[b]{0.48\textwidth}
		\centering
		\includegraphics[width=\textwidth]{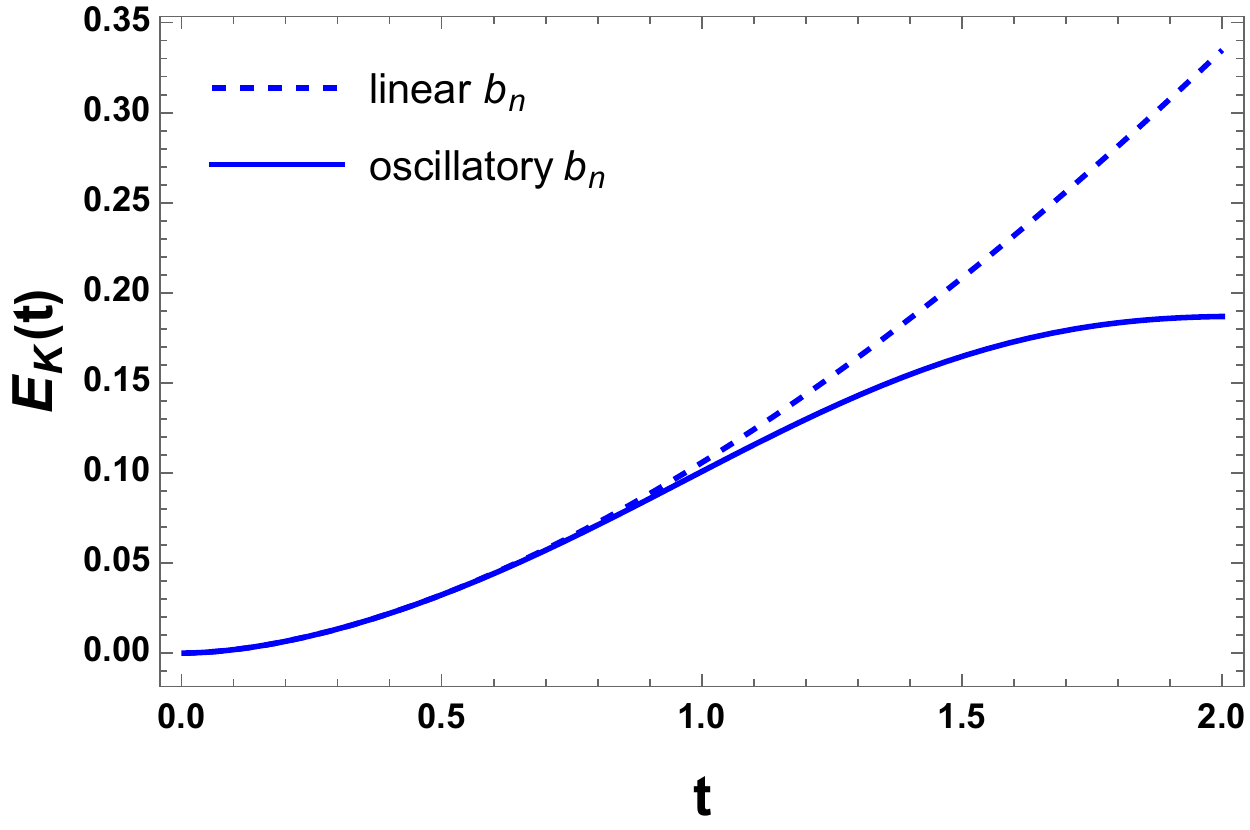}
			\caption{For $q = 100$.}
			\label{fig:var3}
		\end{subfigure}
		\hfill
		\begin{subfigure}[b]{0.48\textwidth}
		\centering
		\includegraphics[width=\textwidth]{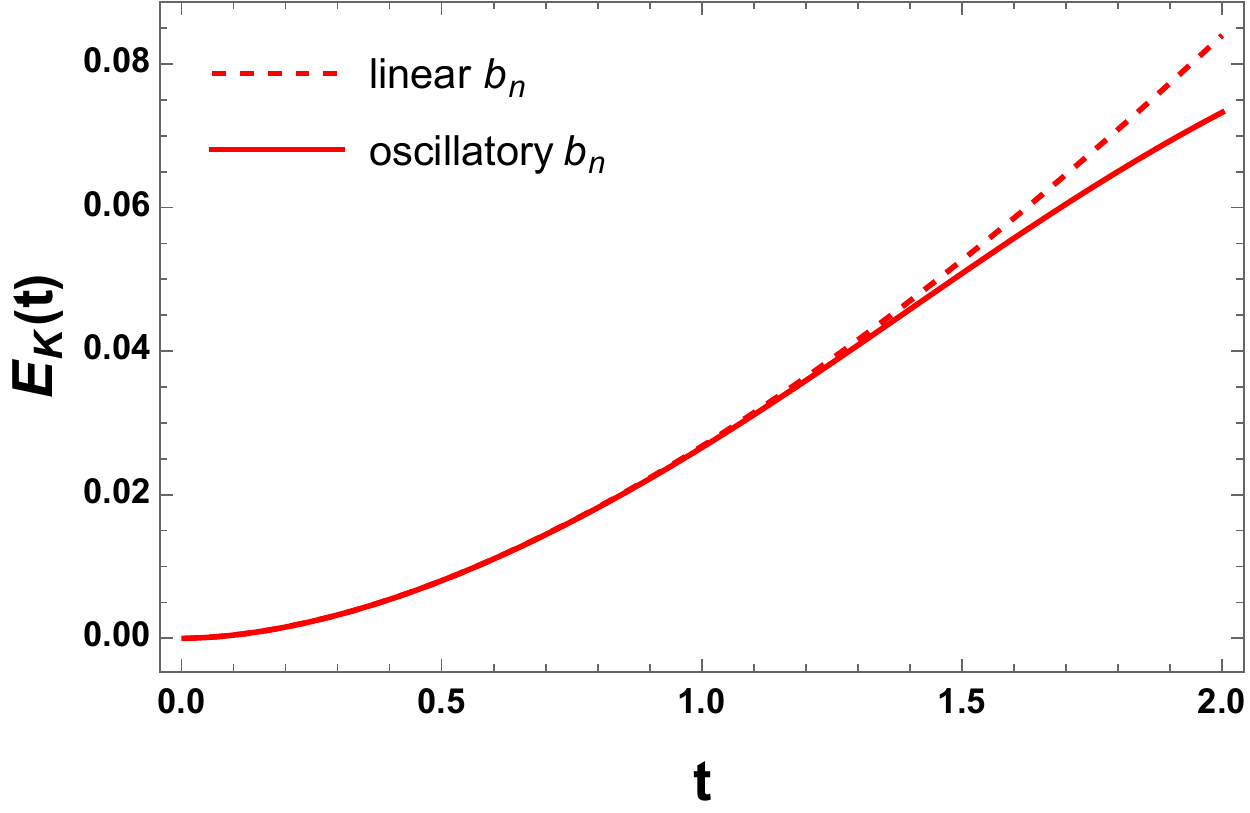}
			\caption{For $q = 500$.}
			\label{fig:var4}
		\end{subfigure}
		\caption{The behavior of K-entropy  is shown for (a) $q=100$ and (b) $q = 500$ respectively, with the Lanczos coefficients \eqref{bnsubleading}. The dashed line demonstrates the analytic result \eqref{cent}.}
		\label{entplots}
\end{figure}

\begin{figure}[t]
		\centering
		\begin{subfigure}[b]{0.48\textwidth}
		\centering
		\includegraphics[width=\textwidth]{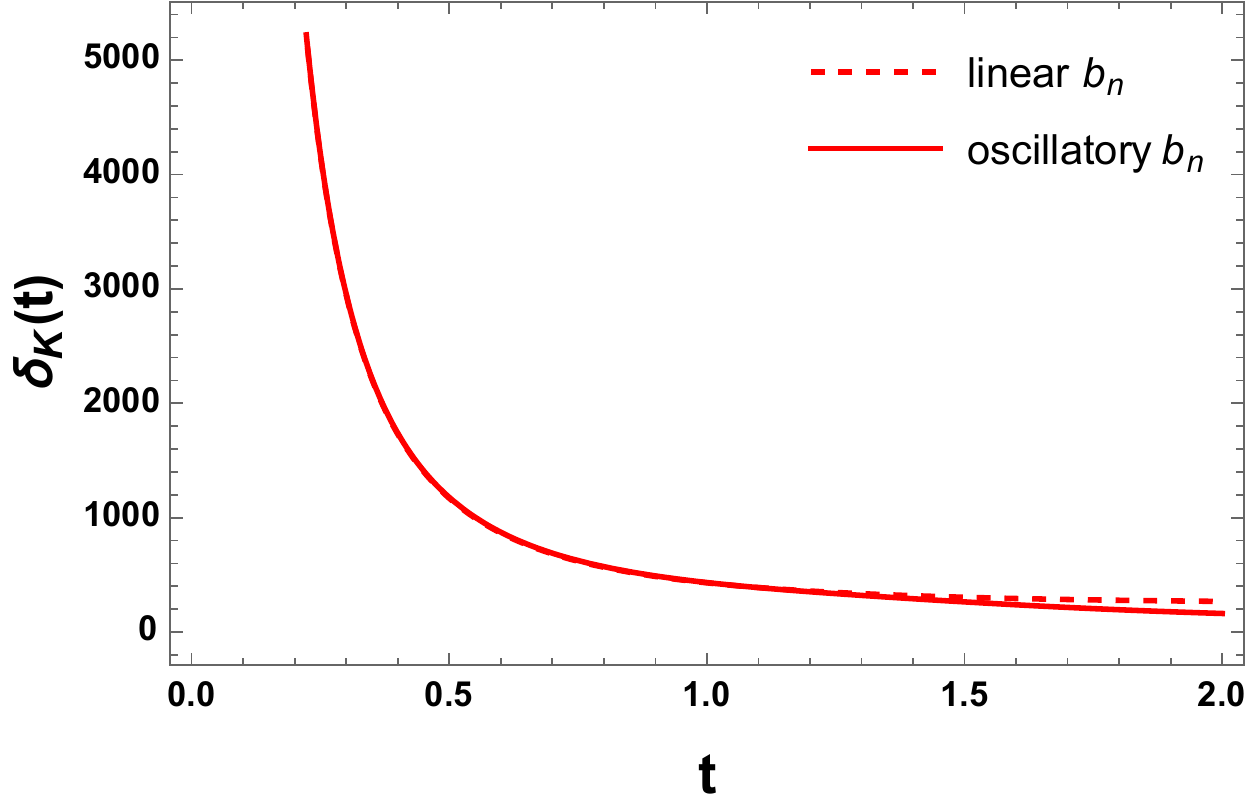}
			\caption{For $q = 500$.}
			\label{fig:sk3}
		\end{subfigure}
		\hfill
		\begin{subfigure}[b]{0.48\textwidth}
		\centering
		\includegraphics[width=\textwidth]{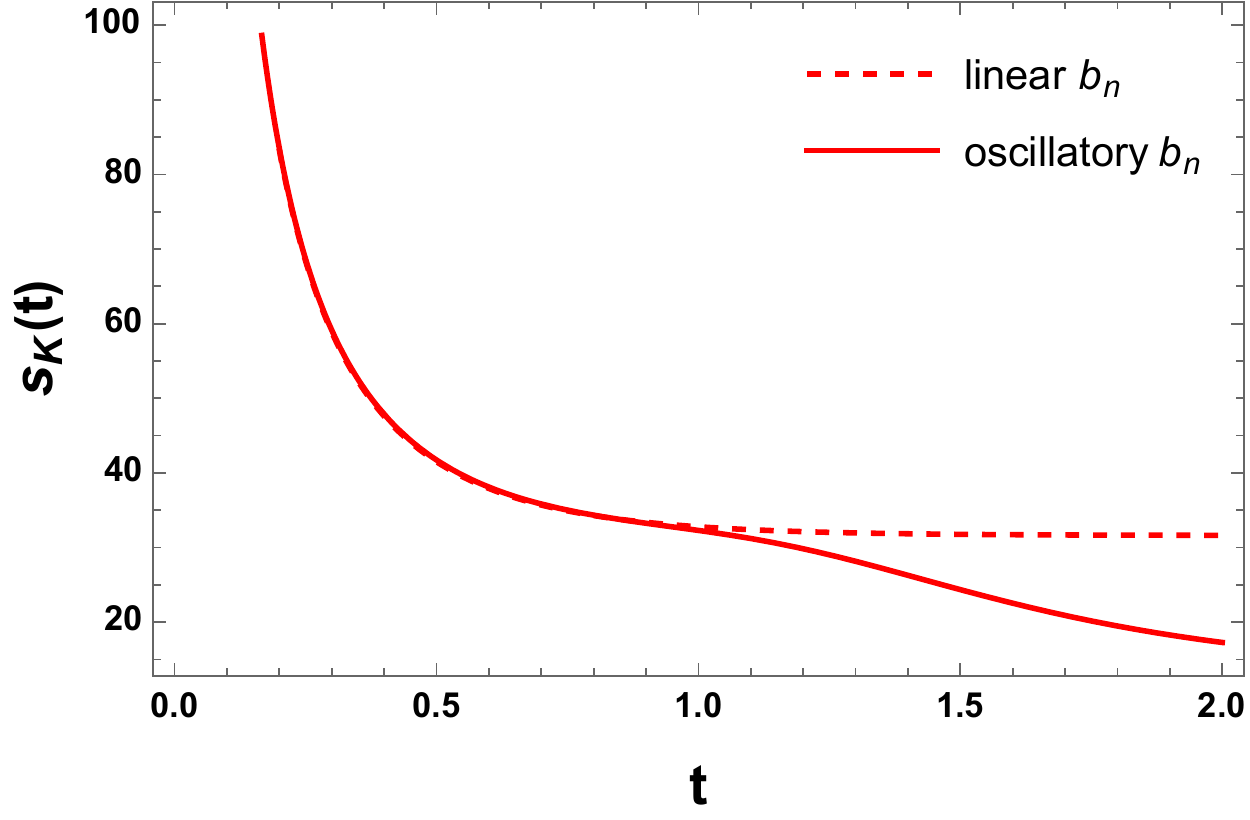}
			\caption{For $q = 500$.}
			\label{fig:sk4}
		\end{subfigure}
		\caption{The behavior of K-variance and K-skewness is shown for $q = 500$ with the Lanczos coefficients \eqref{bnsubleading}. The dashed lines demonstrate the analytic result \eqref{cc} and \eqref{skr} respectively.}
		\label{vrskplots}
\end{figure}
In Fig.\,\ref{fig:kc3}, we demonstrate the behavior of the complexity. A crucial point that the above picture incorporates is the truncation of complexity at $1/q^2$ order. This, along with the finite truncation with a limited number of coefficients makes the $1/q^2$ result grow faster than the $1/q$ result. As a perturbative series in $1/q$, the K-complexity contains orders higher than $1/q$. It is however important to mention that in doing this, we miss some contribution to those higher orders (for a fixed $n$) that arise from higher order corrections in the auto-correlation function (which we have restricted to $\mathcal{O}(1/q^2)$). Therefore, the situation observed in Fig.\,\ref{fig:kc3} is an artifact of our treatment of K-complexity as a perturbative series in $1/q$. Such perturbative effects were not present in previous studies of systems with oscillatory Lanczos coefficients.

\subsection{Krylov cumulants without truncation in $q$} \label{witr}

Several previous studies show that the oscillatory Lanczos coefficients make the complexity growth slower, see \cite{Rabinovici:2021qqt, Rabinovici:2022beu, Trigueros:2021rwj} for examples.\footnote{We thank Anatoly Dymarsky and the anonymous referee for pointing this out to us.} We see that this is indeed true once we get rid of the truncation in $q$. For this purpose, we take the Lanczos coefficients \eqref{bnsubleading} up to $n = 14$ and numerically solve the recursive differential equation \eqref{recde} as Schr\"odinger equation. The numerical results do not consider any  truncation in $q$. The results of the K-complexity, K-entropy K-variance, and K-skewness are shown in Fig.\,\ref{complots}, Fig.\,\ref{entplots} and Fig.\,\ref{vrskplots}. We indeed see that the oscillatory behavior of $b_n$ reduces the complexity, entropy, and all other cumulants compared to the results for linear $b_n$. Moreover, increasing $q$ tends to be closer to the analytic results due to $\mathcal{O}(1/q^2)$ suppression. This also supports the fact that oscillatory Lanczos coefficients decrease the growth of Krylov cumulants, provided we compute them exactly without any truncation in $q$.

We briefly discuss the inexact nature of both these methods. Due to an increase in computational difficulty with $n$, we are only able to capture the first few Lanczos coefficients (Appendix \ref{appa}).\footnote{We again stress that both the truncated $q$ and without $q$ truncated result contains the ``truncated'' $n$ effect. The truncation in $q$ is a different choice than ``$n$-truncation''. When we truncate in $q$ (sec 4.1, truncation in $n$ is still present) we consider the expression for the Krylov cumulants to order $1/q^2$ specifically, in terms of a perturbative series. However, when we choose not to truncate in $q$ (sec 4.2, again truncation in $n$ is present), we essentially do not perform any perturbation series in $q$, and therefore include all contributions in the Krylov cumulants.} Therefore, we lose some information regarding the rest of the Krylov space. This is also reflected in the definition of the auto-correlation function, which in principle, contains all the information of the Lanczos coefficients. In other words, a full set of Lanczos coefficients is required to reconstruct the exact auto-correlation function. 
However, the way in which this lack of information reflects in the two methods is different. The analytical method, in which we truncate in $q$ relies on using the recursive Schr\"odinger equations \eqref{recde} to determine the exact analytical expressions for each wavefunction $\varphi_{n}$ using the previous $n - 1$ wavefunctions. We are able to do this up to $n = 13$ since we have access to $14$ Lanczos coefficients. The loss of information here is captured by the probability $\sum_{n}|\varphi_{n}(t)|^{2}$ which significantly drops from unity after some time. Any results after that time are not reliable as the initial operator has spread outside the Krylov subspace described by the $13$ $\varphi_{n}$'s which we know. The second method is conceptually different. Again, we start with the $14$ Lanczos coefficients we have. However, this method effectively assumes this finite set of Lanczos coefficients to describe the full Krylov space (by imposing probability conservation). This happens because this method \textit{solves} the recursive Schr\"odinger equations \eqref{recde} to determine the Krylov wavefunctions. Therefore, the results obtained for the wavefunctions (and by extension, the rest of the Krylov cumulants) are correct only at small times. At late times, it implies modification of the wavefunctions in such a way that the oscillating coefficients always lead to lower complexity, as shown in Fig.\,\ref{complots}-Fig.\,\ref{vrskplots}. Due to the numerical nature of this method, it is not possible to truncate the wavefunctions or cumulants in $q$. So this method can be useful to determine the effects arising due to truncation in $q$, by comparing it with the results obtained by the truncation in $q$.

\section{Complexity and hyperfast scrambling in $\mathrm{DSSYK}_{\infty}$}\label{dShyperfast}
\label{sec:hyper}
In this section, we consider a special limit, in contrast to previous sections, where we considered the two-stage limit. Concisely, in this limit, $q$ is held fixed while we take $N \rightarrow \infty$. Here we would like to scale $q$ as we scale $N \rightarrow \infty$. We particularly take $q \sim \sqrt{N}$, such that $\lambda = q^2/N$ is held fixed. This limit is known as the double-scaled limit. Here we should point out that previous studies \cite{Susskind:2021esx, Lin:2022nss} proposed a more generalized scaling $q \sim N^p$. However, here we consider $p = 1/2$, which is consistent with the semi-classical limit and the existence of the separation of scales \cite{Susskind:2022bia}.\footnote{We thank Leonard Susskind for clarifying this to us.} Due to such scaling, the $k$-locality is not preserved. Further, we focus on the infinite (Boltzmann) temperature limit (see later) and call it $\mathrm{DSSYK}_{\infty}$. In this limit, the Hamiltonian can be written as
\begin{align}
    H = q \, i^{q/2} \sum_{1 \leq i_1 < i_2 \cdots  i_q \leq N} \mathbf{j}_{i_1 \cdots i_q} \, \chi_{i_1} \cdots \chi_{i_q}
    = i^{q/2} \sum_{1 \leq i_1 < i_2 \cdots  i_q \leq N} \tilde{j}_{i_1 \cdots i_q} \, \chi_{i_1} \cdots \chi_{i_q}\,, \label{hamr}
\end{align}
where $\tilde{j}_{i_1 \cdots i_q} = q  \, \mathbf{j}_{i_1 \cdots i_q}$. To obtain the Hamiltonian, we have multiplied the Hamiltonian \eqref{sykh1} by $q$. The fields satisfy the anti-commutation relation $\{\chi_i, \chi_j\} = 2\delta_{ij}$, as usual. In other words, we can think \eqref{hamr} as the given Hamiltonian with the variance of the distribution of the random fields in \eqref{hamr} is
\begin{align}
    \braket{\tilde{j}^2_{i_1 \cdots i_q}} = q^2 \braket{\mathbf{j}^2_{i_1 \cdots i_q}}  = \frac{q!}{2 N^{q-1}} \mathcal{J}^2 \,. \label{varnew}
\end{align}
This rescaling of $q$ amounts to bringing an inverse factor of $q$ in the rescaling of time. Hence, the time coordinate has to be replaced by $q t$ \cite{Lin:2022nss}. In other words, this is nothing but the relation between string time $\tilde{t}_s$ and cosmic time $\tilde{t}_c$, with $\tilde{t}_s = q \tilde{t}_c$ \cite{Susskind:2022bia, Rahman:2022jsf}. This rescaling is absolutely vital to keep the Hamiltonian and the associative quantities finite in the double-scaled limit.

\subsection{Hyperfast scrambling: Lanczos coefficients and Krylov complexity}

Now, we would like to understand the scrambling behavior governed by the Hamiltonian \eqref{hamr}. In \cite{Lin:2022nss, Susskind:2022bia, Rahman:2022jsf}, the scrambling is observed by using a classic epidemic model \cite{Qi:2018bje}. Here wish to study the scrambling by a more refined and systematic probe, the Krylov complexity. However, we should mention that here we follow \cite{Susskind:2022bia}, and our results are at the ``zeroth'' level. A more systematic study involving the chord diagrams \cite{Berkooz:2018jqr} will be reported elsewhere.

Given the Hamiltonian \eqref{hamr} with the distribution \eqref{varnew}, we can ask how an initial operator evolves under the time-evolution by \eqref{hamr}. We start with the normalized initial operator $\chi(0)$. The auto-correlation function is given by the two-point function $\mathcal{C} (t_c) = \braket{\chi(\tilde{t}_c) \chi(0)}$ with respect to the infinite-temperature inner product, where $\tilde{t}_c$ denotes the cosmic time \cite{Susskind:2022bia}. This has been computed in \cite{Lin:2022nss}. It is given by
\begin{align}
    \mathcal{C}(\tilde{t}_c) = [\sech (q \mathcal{J} \tilde{t}_c)]^{2/q}\,.
\end{align}
It is important to note that this is not the autocorrelation function for the DSSYK$_{\infty}$, and consequently not the bulk propagator for de Sitter geometry. That is because, in the true double-scaled limit, one must have $q \rightarrow \infty$. However, we chose to retain $q$ explicitly (can be replaced by $\sqrt{\lambda N}$, where $\lambda$ is an $\mathcal{O}(1)$ number and $N \rightarrow \infty$) so that we are able to study the exact nature the Lanczos coefficients and the Krylov moments as $q \rightarrow \infty$. Importantly, the autocorrelation function for DSSYK$_\infty$ (in the $(q,N)$ language) is recovered in the $q \rightarrow \infty$ limit \cite{Susskind:2022bia, Lin:2022nss}. Consequently, the true ``hyperfast scrambling'' nature is revealed at the $q \rightarrow \infty$ limit of the Lanczos coefficients and Krylov cumulants. This will be demonstrated with explicit calculations in the rest of this section.

The moments are computed using \eqref{momentsdef}. They are given by
\begin{align}
    m_{2n} = q^{2n-1} \mathcal{J}^{2n} T_{n-1} + O(q^{2n-2})\,, ~~~~~~ n \geq 1\,.
\end{align}
where $\{T_{n-1}\}_{n=1}^{\infty} = \{1, 2, 16, 272, 7936, \cdots \}$ are the Tangent numbers. Note that here a multiplicative $q^{2n-1}$ factor appears compared to $1/q$ factor in Eq.\eqref{mom1}, which can also be understood as scaling $\mathcal{J} \rightarrow q \mathcal{J}$ in Eq.\eqref{mom1}. This makes the growth of the moments quite rapid as $q \rightarrow \infty$ in large $n$. The Lanczos coefficients are calculated from the recursive algorithm \eqref{mombn} as
\begin{align}
b_n = \mathcal{J} \sqrt{n q (2 + (n-1) q)}\,, ~~~~~ n \geq 1.
\end{align}
We write it in a more suggestive form 
\begin{align}
    b_n = \alpha \sqrt{n (n - 1 + \eta)}\,, ~~~~~ \alpha = q \mathcal{J}\,, ~~~ \eta  = \frac{2}{q}\,.
\end{align}
The generic expressions of $b_n$ can be compared with \cite{Parker:2018yvk}. Especially, we see that $b_n$'s diverge as $q \rightarrow \infty$. This apparently violates the statement of the universal operator growth hypothesis \cite{Parker:2018yvk}, which states that $b_n$ cannot show more than linear growth in $n$. However, the hypothesis is based on locality, which is violated in the double-scaled limit. Hence, the diverging Lanczos coefficients do not contradict the hypothesis. We further observe that
\begin{align*}
     \int_{1}^{q^2+1}\frac{\mathrm{d} x}{\sqrt{x(x-1 + 2/q)}}< \sum_{n = 1}^{q^2}  \frac{1}{\sqrt{n(n-1 + 2/q)} } < \sqrt{\frac{q}{2}} +\int_{1}^{q^2}\frac{\mathrm{d} x}{\sqrt{x(x-1 + 2/q)}}\,.
\end{align*}
Multiplying the above inequality by $1/q$ and taking the limit $q \rightarrow \infty$, we get
\begin{align}
 0 < \lim_{q \rightarrow \infty}\frac{1}{q}\sum_{n = 1}^{q^2}  \frac{1}{\sqrt{n(n-1 + \eta)} } <  
0\,.
\end{align}
Using the Squeeze theorem, we see that the middle term of the above inequality also evaluates to zero. Thus, in general, we have
\begin{align}
    \lim_{N \rightarrow \infty} \sum_{n = 1}^N \frac{1}{b_n} =  \frac{1}{q \mathcal{J}} \lim_{N \rightarrow \infty} \sum_{n = 1}^N  \frac{1}{\sqrt{n(n-1 + \eta)} } = \mathrm{finite} \,,
\end{align}
with $\eta = 2/q$. Hence, Carleman’s condition \cite{carleman} is not satisfied, and the Hamburger moment problem could be indeterminate. However, we should understand that Carleman’s condition is not a necessary condition (but it is a sufficient condition for determinacy), and hence the convergent result does not necessarily make the problem indeterminate, in general.\footnote{One can, however, consider Krein’s condition as a sufficient condition for the indeterminacy \cite{momentlecture}. We have not considered this in this paper.} If the problem is indeed indeterminate, then it is an interesting open question to see whether this is linked to the absence of locality in the Hamiltonian.

Furthermore, the Krylov basis functions can be obtained by solving the differential equation \eqref{recde}. They read
\begin{align}
    \varphi_{n}(\tilde{t}_c) = \sqrt{\frac{(\eta)_n}{n!}} \tanh^n (\alpha \tilde{t}_c) \sech^{\eta} (\alpha \tilde{t}_c)\,, \label{basis}
\end{align}
where $(\eta)_n = \eta (\eta + 1) \cdots (\eta + n -1)$ is the Pochhammer symbol \cite{Parker:2018yvk}. The complexity is given by performing the weighted sum \eqref{kcom} as
\begin{align}
    C_K(\tilde{t}_c) = \eta \sinh^2 (\alpha \tilde{t}_c) =  \frac{2}{q} \sinh^2 (q \mathcal{J} \tilde{t}_c) \sim \frac{2}{q} e^{2 q \mathcal{J} \tilde{t}_c}  \,.
\end{align}
The exponential growth in the numerator dominates the polynomial growth in the denominator. The Krylov complexity grows hyperfast.\footnote{In \cite{Rahman:2022jsf}, an example was given where complexity (computed in the circuit model) is supposed to grow linearly with time, not hyperfast. However, this depends on the ``appropriate'' definition of complexity. For example, the holographic complexity \cite{Jorstad:2022mls} has also shown to be divergent in specific limits.} The Lyapunov exponent is $\lambda_L = q \mathcal{J}$, consistent with the observation in \cite{Lin:2022nss}. Here we should also mention that the hyperfast scrambling is only visible in the cosmic time ($\tilde{t}_c$) unit, but not the string unit ($\tilde{t}_s$). The scrambling time can be found by observing $C_K(\tilde{t}_c^{*}) \sim O(1)$; thus, the scrambling time is
\begin{align}
    \tilde{t}_c^{*} \sim \frac{1}{2 q \mathcal{J}} \ln q\,,
\end{align}
and marked by the (energy) scale $\mathcal{J}$ according to $\tilde{t}_c^{*} \sim \mathcal{J}^{-1}$, as conjectured in \cite{Susskind:2021esx}. The denominator dominates for large $q$. Hence,  scrambling time shrinks to zero ($\tilde{t}_c^{*} \rightarrow 0$) in the double scaling limit, i.e., the scrambling is instantaneous.\footnote{A similar timescale also appears in matrix models \cite{Jafferis:2022uhu}.} This describes the ``hyperfast'' scrambling termed in \cite{Lin:2022nss}. From the dual gravity picture, this seems reasonable as the holographic degrees of freedom live on the horizon of the dS, not in the asymptotic boundary as in AdS \cite{Dong:2018cuv}. This hyperfast scrambling seems to violate the chaos bound \cite{Maldacena:2015waa} at first glance. However, as has been argued \cite{Susskind:2021esx}, the important assumption in the derivation of the chaos bound is the $k$-locality of the Hamiltonian. The double-scaled SYK violates this assumption of the $k$-locality (for which we have got divergent Lanczos coefficients) and thus makes it possible for the model to violate the bound.

For the usual SYK Hamiltonian \eqref{sykh1}, For $\mathcal{C}(\tilde{t}_s) = [\sech (\mathcal{J} \tilde{t}_s)]^{2/q}$, one can directly find $b_n = \mathcal{J} \sqrt{n (n - 1 + \eta)}$ with $\eta  = 2/q$. Here we use the strings unit as we considered the usual SYK. We can directly write the Krylov basis functions
\begin{align}
     \varphi_{n}(\tilde{t}_s) = \sqrt{\frac{\Gamma (n + 2/q)}{n! \, \Gamma (2/q)}} \tanh^n (\mathcal{J} \tilde{t}_s) \sech^{2/q} (\mathcal{J} \tilde{t}_s)\,, \label{bas2}
\end{align}
so that the probability is $ |\varphi_{n}(\tilde{t}_s)|^2$. The probability exactly matches with \eqref{prob1}. Hence the complexity is $C_K(\tilde{t}_s) = (2/q) \sinh^2 (\mathcal{J} \tilde{t}_s) \approx (2/q) \exp(2 \mathcal{J} \tilde{t}_s)$. This is consistent with \eqref{cc}. Note that here we do not have any $q$-dependence in the exponential term, which is required for the hyperfast scrambling \cite{Lin:2022nss}. The scrambling time is $\tilde{t}_s^{*} \sim 1/(2\mathcal{J}) \ln q$,  which does not shrink to zero. This is the usual case for the fast scrambler \cite{Sekino:2008he, Lashkari:2011yi}, where $k$-locality is preserved.

\subsection{Computing higher Krylov cumulants}
In the $\mathrm{DSSYK}_{\infty}$, we move to compute the higher cumulants, especially the variance and the skewness. The Krylov basis functions are given by
\begin{align}
     \varphi_{n}(\tilde{t}_c) = \sqrt{\frac{\Gamma (n + 2/q)}{n! \, \Gamma (2/q)}} \tanh^n (q \mathcal{J} \tilde{t}_c) \sech^{2/q} (q \mathcal{J} \tilde{t}_c)\,,
\end{align}
Note the appearance of the $q$ factor in the argument, which is a crucial difference from \eqref{bas2}. This implies the variance and the skewness are given by
\begin{align}
    \delta_K (\tilde{t}_c) = \frac{q}{2} \coth^2 (q \mathcal{J} \tilde{t}_c)\,, ~~~~~ s_K(\tilde{t}_c) = \sqrt{2 q} \coth(2 q \mathcal{J} \tilde{t}_c)\,.
\end{align}
The expressions are very similar to \eqref{cc} and \eqref{skr} except for the crucial $q$ factor in the argument. However, they are not exponentially diverging like the complexity due to the presence of the ``$\coth$'' term. This is because for large $q$, we have $\coth (q \mathcal{J} \tilde{t}_c) \rightarrow 1$. Hence, in the large $q$ limit, variance, and skewness are proportional to $q$ and $\sqrt{q}$, respectively.

All the expressions can be derived from the probability distribution with a slight modification in the argument
\begin{align}
    P_n(\tilde{t}_c) = \frac{\Gamma (n+2/q)}{\Gamma (n+1) \,\Gamma(2/q)}\, \frac{\tanh^{2 n}(q \mathcal{J} \tilde{t}_c)}{\cosh^{4/q}(q \mathcal{J} \tilde{t}_c)}\,, \label{prob1new}
\end{align}
The probability is conserved for all time i.e., $\sum_{n=0}^{\infty} P_n(\tilde{t}_c) = 1$. In comparison with our previous discussions, this distribution also defines the ``size'' of the operator in the hyperfast scrambling regime.

We briefly point out the apparent connection of hyperfast scrambling in de Sitter (dS) space. It is argued in \cite{Lin:2022nss, Susskind:2022bia, Rahman:2022jsf} that the parameter $\mathcal{J}$ is related to the horizon radius $L_c$ of the dS. They are inversely proportional to each other i.e., $\mathcal{J} = 1/L_c$. The standard Boltzmann temperature in dS is infinite \cite{Susskind:2021esx}. A possible way to think in terms of the finite value of the partition function. As Hamiltonian scales with $q$, one requires the temperature to scale linearly with $q$. Hence, in the double-scaled limit, the temperature becomes infinite. From the gravity side, this happens due to the existence of the flat entanglement spectrum \cite{Dong:2018cuv, Chandrasekaran:2022cip}. However, the effective temperature (remarked as the tomperature in \cite{Susskind:2021esx, Lin:2022nss}) $\mathcal{T}$ is defined as $\mathcal{T} = 2 \mathcal{J} = 2/L_c$, which is independent of $q$. Hence, the complexity grows as
\begin{align}
    C_K(\tilde{t}_c) = \frac{2}{q} e^{q \mathcal{T} \tilde{t}_c}\,.
\end{align}
Thus, even if we are considering the infinite-temperature limit, the growth is controlled by the effective temperature.

\section{Conclusion and outlook}

In this paper, we have considered the SYK$_q$ system of Majorana fermions in the large $q$ and the large $N$ limit. We discuss two distinct limiting procedures and their aspects from the perspective of Krylov complexity. The first limiting procedure we consider is the two-stage limit, where the large $N$ Schwinger-Dyson equations are solved order by order in $1/q$. The first-order corrections were considered in \cite{Parker:2018yvk}, while in this work we extend it to the second-order corrections \cite{Tarnopolsky:2018env} and consider their contribution to Krylov complexity. We have discussed the effects of the $\mathcal{O}(1/q^2)$ term on the Lanczos coefficients, Krylov basis wavefunctions, and their first few cumulants. We find that the consideration of the second order term is valid up to a cutoff time, fixed by the value of $q$. This cutoff is reflected in the behavior of the moments of the Krylov wavefunction probability distribution. We discuss the deterministic nature of this limiting procedure within the purview of the Hamburger moment problem. The second limiting procedure we consider is the double-scaling limit, where both $q$ and $N$ are sent to infinity while holding an appropriately defined ratio of the two constants fixed. In other words, we consider the specific case where $\lim_{\{q, N\} \rightarrow \infty} q /\sqrt{N} = \lambda$. This limit has various interesting implications, including a proposed duality (of SYK$_q$ in this limit) to de-Sitter space. We consider an appropriate scaling of the SYK$_q$ Hamiltonian to obtain a maximally mixed density matrix at infinite temperature as advocated in \cite{Susskind:2022bia}. We then evaluate the Krylov complexity in this model and demonstrate that it exhibits hyperfast scrambling in cosmic timescales. This model is known to violate $k$-locality (which is a central assumption in most discussions about scrambling), reflected via the vanishing scrambling time in the double scaling limit. Finally, we discuss the nature of this limiting procedure with respect to the Hamburger moment problem and find that it could be non-deterministic. 

We conclude the paper with a few interesting future directions. Our results are entirely in the infinite-temperature regime. It is unclear whether this hyperfast growth is still valid in 
the finite-temperature limit. In such a case, we believe that a more systematic understanding from the boundary side is required, particularly in terms of the chord diagrams \cite{Berkooz:2018jqr}, where the auxiliary Hilbert space can be treated as a Krylov-like subspace \cite{Berkooz:2018qkz}. Especially, an interesting direction we hope to return to is the bulk computation for the same, where the bulk Hilbert space is formed by a Krylov-like construction \cite{Lin:2022rbf}. Moreover, the $1/q^2$ correction is important since it might shed light on the contribution of the disconnected geometries \cite{Kar:2021nbm}, which provide the subleading corrections of Lanczos coefficients and the complexity from the gravity side. Finally, it is also interesting to consider other generalizations of SYK, namely its supersymmetric generalization \cite{He:2022ryk} in the large $q$ \cite{Fu:2016vas}, and particularly its double-scaled limit \cite{Berkooz:2020xne}. A naive computation suggests that one might expect two sets of Lanczos coefficients, with the two sets of moments expressed in terms of Secant and Tangent numbers. We hope to address them in the future.

\section*{Acknowledgements}
We would like to thank Pawel Caputa, Anatoly Dymarsky, Chethan Krishnan, Tatsuma Nishioka, Prasanth Raman, Tadashi Takayanagi, and Masataka Watanabe for fruitful discussions and comments on the draft. We especially thank Leonard Susskind for the correspondence regarding \cite{Susskind:2022bia}. We also thank the anonymous referee to point out several subtle issues which improve the quality of the paper. Part of the work was presented (by P.N.) in the Extreme Universe circular meeting. B.B. is supported by the Ministry of Human Resource Development (MHRD), Government of India, through the Prime Ministers' Research Fellowship. The work of P.N. is supported by the JSPS Grant-in-Aid for Transformative Research Areas (A) ``Extreme Universe'' No. 21H05190.

\appendix

\section{Appendix: Computation of the auto-correlation function}
\label{appauto}
In this Appendix, we outline the process of evaluating the integral \eqref{htau}. First, we note down the integral
\begin{align}
    \int_{0}^{\pi v/2} dy \,\ell(y) = - \frac{\pi^2 v^2}{24 \cos^2 (\pi v/2)} \big[\pi v + 3 \sin (\pi v) \big]\,,
\end{align}
which has the following property $\int_{-\pi v/2}^{0} dy \,\ell(y) = \int_{0}^{\pi v/2} dy \,\ell(y)$.
However, more generically, we need to solve a more difficult and non-trivial integral of the form
\begin{align}
    I(x) = \int_{0}^{x}dy\, \ell(y)\,.
\end{align}
It is important to note that the integrand $\ell(y)$ contains $v$, which we require to be vanishing when working in the infinite-temperature limit. Therefore, the form of the integrand we shall use is the following
\begin{align}
    \ell(y) = 2 \ln (\sec y) - \cos^{2}y \, \mathrm{Li}_2(-\tan^{2}y)\,.
\end{align}
We split the integral into two parts. The first integral is easy to evaluate. It gives
\begin{align}
    \int dy\, 2\ln(\sec y) = 2 \left[-\frac{i}{2}  \mathrm{Li}_2\left(-e^{2 i y}\right)-\frac{i y^2}{2}+y \log \left(1+e^{2 i y}\right)+y \log (\sec y)\right]\,. \label{int1fin}
\end{align}
Evaluating the second integral is non-trivial. We proceed as follows. We make a change of variable $z = \tan y$, and cast the integral in terms of $z$ as
\begin{align}
    \int dy\, \cos^{2}y\, \mathrm{Li}_{2}(-\tan^{2}y)  = \int dz\, (1 + z^2)^{-2} \, \mathrm{Li}_{2}(-z^{2}) \,.
\end{align}
The integral contour can be rotated by the following wick rotation $z = i s$, where the integral is now known. This gives
\begin{align}
    &i\int ds\, (1 - s^2)^{-2}\, \mathrm{Li}_{2}(s^{2}) = \frac{i}{12}  \Bigg[6 \text{Li}_2\left(s^2\right) \Big(\tanh^{-1}s-\frac{s}{s^2-1}\Big)+6 \text{Li}_2\Big(\frac{1-s}{2}\Big) \notag \\ &-6 \text{Li}_2\Big(\frac{s+1}{2}\Big)+12 \text{Li}_3(1-s)-12 \text{Li}_3(s+1)+12 \text{Li}_2(s+1) \log (s+1) \notag \\ &+12 \text{Li}_2(s) \log (1-s)+3 (2 \log (-s)+1) \log^2(s+1) \notag\\ &+\log (1-s) \left(\log (1-s) (6 \log s-3)-2 \pi^2\right) + 12 \log 2 \tanh ^{-1}s\Bigg]\,.
\end{align}
We need to reverse the coordinate transformations. We replace $s = - i \tan y$. This gives us
\begin{align}
    &\int dy \,\cos^{2}y \mathrm{Li}_{2}(-\tan^{2}y) = -\frac{i}{2} \text{Li}_2\Big(\frac{1}{1+e^{2 i y}}\Big)+\frac{y}{2} \text{Li}_2\left(-\tan^2 y\right)+\frac{i}{2}  \text{Li}_2\Big(\frac{1}{2} (i \tan y +1)\Big) \notag \\ &+i \text{Li}_3(i \tan y +1)-i \text{Li}_3(1-i \tan y )+i \text{Li}_2(-i \tan y) \log (1+i \tan y) \notag \\ &+i \text{Li}_2(1-i \tan y) \log (1-i \tan y)+\frac{1}{4} \sin 2y\, \text{Li}_2\left(-\tan^2 y\right)+ y \log 2 \notag \\ &-\frac{i}{4} \log ^2(1+i \tan y)+\frac{i}{2}  \log (-i \tan y) \log^2(1+i \tan y)+\frac{i}{4}  \log ^2(1-i \tan y)\notag\\ &+\frac{i}{2} \log (i \tan y) \log^2(1-i \tan y)-\frac{i \pi^2}{6}  \log (1+i \tan y)\,. \label{int2fin}
\end{align}
The limits of this integral are $0$ and $x$. The lower limit evaluates to zero, so the only non-zero contribution comes from the upper limit. We simply replace $y$ by $x$ in \eqref{int2fin}. The upper limit in \eqref{int1fin} is likewise a replacement of $y$ by $x$, while the lower limit evaluates to $i \pi^{2}/12$. Hence the full expression for the integral $I(x)$ is
\begin{align}
    \int_{0}^{x}dy\,\ell(y) &= 2 \left[-\frac{i}{2}  \text{Li}_2\left(-e^{2 i x}\right)-\frac{i x^2}{2}+x \log \left(1+e^{2 i x}\right)+x \log (\sec x)\right] \notag \\
    &\frac{i}{2} \text{Li}_2\Big(\frac{1}{1+e^{2 i x}}\Big)-\frac{x}{2} \text{Li}_2\left(-\tan^2 x \right)-\frac{i}{2} \text{Li}_2\Big(\frac{1}{2} (i \tan x +1)\Big)\notag \\ &-i \text{Li}_3(i \tan x +1)+i \text{Li}_3(1-i \tan x )-i \text{Li}_2(-i \tan x ) \log (1+i \tan x )\notag \\ &-i \text{Li}_2(1-i \tan x) \log (1-i \tan x)-\frac{1}{4} \sin 2x \, \text{Li}_2\left(-\tan^2 x\right)- x \log 2 \notag \\ &+\frac{i}{4}  \log^2(1+i \tan x)-\frac{i}{2}  \log (-i \tan x) \log^2(1+i \tan x)-\frac{i}{4}  \log^2(1-i \tan x) \notag \\& -\frac{i}{2} \log^2(1-i \tan x) \log (i \tan x)+\frac{i \pi^2}{6}   \log (1+i \tan x) - \frac{i \pi^{2}}{12}\,. \label{intfin}
\end{align}
This is the result that also holds when $v = 0$ in $\ell(y)$. For this case of $v = 0$, we have $x = - \mathcal{J} \tau$. Transforming to the Lorentzian time coordinates, we must now replace $\tau = i t$. Hence in \eqref{intfin} we must insert $x = - i \mathcal{J} t$.  This gives the following expression
\begin{align*}
    &\int_{0}^{-i \mathcal{J} t}dy\,\ell(y) =\frac{i}{12}  \Bigg[-12 \text{Li}_2\left(-e^{2 \mathcal{J} t}\right)+6 \text{Li}_2\Big(\frac{1}{1+e^{2 \mathcal{J} t}}\Big) -6 \text{Li}_2\Big(\frac{1}{2} (\tanh \mathcal{J} t +1)\Big) \notag \\&+12 \text{Li}_3(1-\tanh \mathcal{J} t)-12 \text{Li}_3(\tanh \mathcal{J} t +1) +12 \text{Li}_2(\tanh \mathcal{J} t) (\log (1-\tanh \mathcal{J} t) \notag \\ &+\log (\tanh \mathcal{J} t+1)) +4 \pi^2 \mathcal{J} t+(6 \log (\tanh (\mathcal{J} t))-3) \log^2(1-\tanh \mathcal{J} t) -\pi^2 \notag \\ &+3 \text{Li}_2\left(\tanh^2 \mathcal{J} t\right) (2 \mathcal{J} t+\sinh 2 \mathcal{J} t -2 \log (\tanh \mathcal{J} t +1))+12 \mathcal{J} t (\mathcal{J} t+\log 2) \notag \\ & +3 (1-2 \log (-\tanh \mathcal{J} t)) \log^2(\tanh \mathcal{J} t +1) -24 \mathcal{J} t \left(\log \left(e^{2 \mathcal{J} t}+1\right)+\log (\text{sech} \mathcal{J} t )\right)\Bigg]\,.
\end{align*}
All of the expressions can be plugged into $h(\tau)$, and thus we get an exact closed-form expression of the auto-correlation function up to $O(1/q^2)$. 
\section{Appendix: Moments and Lanczos coefficients in the subleading order} \label{appa}
Here, we list down the exact expressions of the first 30 moments and 14 Lanczos coefficients
\begin{align}
    m_2 &= \frac{\mathcal{J}^2}{q}\,, ~~ m_4 = \frac{T_1 \mathcal{J}^4}{q} + \frac{64\,\mathcal{J}^4}{q^2} + O(1/q^3) \,, ~~ m_6 = \frac{T_2 \mathcal{J}^6}{q} + \frac{368\,\mathcal{J}^6}{q^2} + O(1/q^3) \,,  \nonumber \\
    m_8 &= \frac{T_3\mathcal{J}^8}{q} + \frac{11440\,\mathcal{J}^8}{q^2} + O(1/q^3)  \,, ~~ m_{10} = \frac{T_4\mathcal{J}^{10}}{q} + \frac{406864\,\mathcal{J}^{10}}{q^2} + O(1/q^3) \,, \nonumber \\ 
    m_{12} &= \frac{T_5 \mathcal{J}^{12}}{q}+ \frac{22093568\, \mathcal{J}^{12}}{q^2} + O(1/q^3)  \,, \nonumber \\
    m_{14} &= \frac{T_6 \mathcal{J}^{14}}{q} +\frac{1640452864\, \mathcal{J}^{14}}{q^2}+ O(1/q^3) \,, \nonumber \\
    m_{16} &= \frac{T_7 \mathcal{J}^{16}}{q} + \frac{160320562176\, \mathcal{J}^{16}}{q^2}+O(1/q^3)\,, \nonumber \\ m_{18} &= \frac{T_8 \mathcal{J}^{18}}{q} +\frac{19948238367744\, \mathcal{J}^{18}}{q^2}+ O(1/q^3) \,, \nonumber \\
    m_{20} &= \frac{T_9 \mathcal{J}^{20}}{q} +\frac{3079484621033472\, \mathcal{J}^{20}}{q^2}+ O(1/q^3) \,,   \nonumber \\ m_{22} &= \frac{T_{10} \mathcal{J}^{22}}{q} + \frac{577609971646545920\, \mathcal{J}^{22}}{q^2}+O(1/q^3) \,, \nonumber \\
    m_{24} &= \frac{T_{11} \mathcal{J}^{24}}{q} +\frac{129388545790244552704\, \mathcal{J}^{24}}{q^2} + O(1/q^3)\,, \nonumber \\
    m_{26} &= \frac{T_{12} \mathcal{J}^{26}}{q}+ \frac{34118755073527150542848\, \mathcal{J}^{26}}{q^2}  + O(1/q^3)\,, \nonumber\\
    m_{28} &= \frac{T_{13} \mathcal{J}^{28}}{q} +\frac{10461593758218426027868160\, \mathcal{J}^{28}}{q^2}+ O(1/q^3)\,, \nonumber\\
    m_{30} &= \frac{T_{14} \mathcal{J}^{30}}{q}+ \frac{3690834649672509819688321024\, \mathcal{J}^{30}}{q^2}  + O(1/q^3)\,.
\end{align}
Here $T_n$'s are the Tangent numbers defined as
\begin{align}
    T_n = \frac{2^{2 n} \left(2^{2 n}-1\right) \left| B_{2 n}\right| }{2 n}\,,
\end{align}
where $B_n$'s are Bernoulli numbers. The Lanczos coefficients are given by
\begin{align}
    b_1 &= \mathcal{J}\sqrt{\frac{2}{q}}\,, ~~ b_2 = \sqrt{2} \mathcal{J} + \frac{31}{\sqrt{2}}  \frac{\mathcal{J}}{q} + O(1/q^2) \,, ~~ b_3 = \sqrt{6} \mathcal{J} - \frac{65}{\sqrt{6}} \frac{\mathcal{J}}{q} + O(1/q^2) \,,  \nonumber \\
    b_4 &= \sqrt{12} \mathcal{J} + \frac{343}{\sqrt{12}} \frac{\mathcal{J}}{q}+ O(1/q^2) \,, ~~ b_5 =\sqrt{20} \mathcal{J} - \frac{8677}{18 \sqrt{20}} \frac{\mathcal{J}}{q} + O(1/q^2)\,, \nonumber \\
     b_6 &= \sqrt{30} \mathcal{J} + \frac{74987}{60\sqrt{30}} \frac{\mathcal{J}}{q}+ O(1/q^2)\,, ~~ b_7 = \sqrt{42} \mathcal{J} - \frac{18811}{12 \sqrt{42}} \frac{\mathcal{J}}{q}+ O(1/q^2)\,, \nonumber \\
     ~~ b_8 &=\sqrt{56} \mathcal{J} + \frac{4830986}{1575 \sqrt{56}} \frac{\mathcal{J}}{q}+ O(1/q^2)\,, ~~ b_9 = \sqrt{72} \mathcal{J} - \frac{17822817}{4900\sqrt{72}} \frac{\mathcal{J}}{q}+ O(1/q^2)\,, \nonumber \\
    b_{10} &= \sqrt{90} \mathcal{J} + \frac{71870293}{11760\sqrt{90}} \frac{\mathcal{J}}{q} + O(1/q^2)\,,~~ b_{11} = \sqrt{110} \mathcal{J} - \frac{2224869499}{317520\sqrt{110}} \frac{\mathcal{J}}{q}+ O(1/q^2)\,,\nonumber\\
    b_{12} &= 2 \sqrt{33} \mathcal{J} +\frac{31137647687}{5821200 \sqrt{33}} \frac{\mathcal{J}}{q} + O(1/q^2)\,,~~ b_{13} = 2 \sqrt{39} \mathcal{J}-\frac{1567183757 }{261360 \sqrt{39}} \frac{\mathcal{J}}{q}+ O(1/q^2)\,,\nonumber\\
    b_{14}&= \sqrt{182} \mathcal{J}+ \frac{555986163137}{32432400 \sqrt{182}} \frac{\mathcal{J}}{q}+ O(1/q^2).
    \label{bnlist}
\end{align}

\bibliographystyle{JHEP}
\bibliography{references}  
\end{document}